\documentclass[11pt]{article}
\usepackage{amsmath,amssymb}
\usepackage{epsfig,graphics,graphicx}
\usepackage{color,multicol}
\usepackage{setspace}
\usepackage{natbib}
\newcommand{\bm}[1]{\boldsymbol{#1}}

\begin{document}

\doublespacing

\title{A Bayesian Nonparametric Markovian Model for Nonstationary Time Series}
\author{Maria DeYoreo and Athanasios Kottas
\thanks{
M. DeYoreo (maria.deyoreo@stat.duke.edu) is Postdoctoral Researcher, Department 
of Statistical Science, Duke University, and A. Kottas
(thanos@ams.ucsc.edu) is Professor of Statistics, Department of Applied Mathematics
and Statistics, University of California, Santa Cruz.}}

\date{}
\maketitle

\begin{abstract}
\noindent
Stationary time series models built from parametric distributions are, in general, limited in scope due 
to the assumptions imposed on the residual distribution and autoregression relationship. We present
a modeling approach for univariate time series data, which makes no assumptions of stationarity, and 
can accommodate complex dynamics and capture non-standard distributions. 
{The model for the transition density arises from the conditional distribution 
implied by a Bayesian nonparametric mixture of bivariate normals.} 
This results in a flexible autoregressive form for the conditional transition density, 
defining a time-homogeneous, nonstationary Markovian model for real-valued data indexed in 
discrete time. To obtain a computationally tractable algorithm for posterior inference, we
utilize a square-root-free Cholesky decomposition of the mixture kernel covariance matrix.
Results from simulated data suggest the model is able to recover challenging transition densities
and nonlinear dynamic relationships. We also illustrate the model on time intervals between 
eruptions of the Old Faithful geyser. Extensions to accommodate higher order structure and 
to develop a state-space model are also discussed.
\end{abstract}


\newpage

\section{Introduction}
\label{sec:intro}

Consider a time series of continuous random variables $(Z_1,\dots,Z_n)$ observed at equally spaced time points $t=1,\dots,n$.  It is common to assume dependence on lagged terms, or that $Z_t$ depends on 
$(Z_{t-1},\dots,Z_{t-p})$, for some $p\geq1$. The relationship between $Z_t$ and $(Z_{t-1},\dots,Z_{t-p})$ 
is generally assumed to be linear, with error terms arising from a given parametric distribution. 
The simplest scenario involves $p=1$ and normally distributed errors, referred to as a first-order 
Gaussian autoregression.

{Time series are generally assumed to be time-homogeneous, that is, the 
transition density that defines the conditional distribution of $Z_t$ given $(Z_{t-1},\dots,Z_{t-p})$
does not change with time. A stronger assumption is that of stationarity, which requires that 
the finite dimensional distributions of the time series are invariant under time shifts.
Weak stationarity requires only the mean to be constant across time and the covariance 
function to be invariant under time shifts.}

{Stationary time series models are not 
appropriate for many applications. Stochastic systems may go through structural changes, 
and as a consequence, the data they produce may require models which change across time. 
While stationarity is a convenient property, stationary models do not allow for this type of 
evolution, as they assume constant means and variances across time. For instance, 
economic time series are commonly believed to be nonstationary \citep[e.g.,][]{fruwirth}.}

{Additionally, customary parametric time series models (both stationary and 
nonstationary) are generally restrictive in terms of the transition and marginal densities they imply. 
Parametric stationary densities are unable to accommodate time series that exhibit asymmetric or 
non-standard marginal distributions. \citet{tong1990} gives an example of a real time series 
that possesses a bimodal marginal distribution. Conditional distributions may also be 
multimodal, for instance when the stock-market is volatile, price changes may be more 
likely to be large in magnitude than near zero, hence it is reasonable to expect a bimodal 
distribution \citep{wongli}.}

Various parametric models have been developed to capture nonlinear autoregressive (AR) 
behavior and/or relax the stationarity assumption. Time-varying autoregressions (TVAR) naturally extend 
AR models, by allowing the parameters to evolve in time, and thus can be used to describe nonstationary 
time series. TVAR models have a dynamic linear model (DLM) representation and belong to the larger class 
of Markovian state-space models. Such models require specification of an observation density and a state 
evolution density, which need not rely on normality or linearity, though these are common assumptions. 

The DLM framework can be made more flexible by combining multiple DLMs, referred to as multiprocess models \citep{westharrison}. Mixture models of various forms have been used to move away from parametric assumptions, and capture changes over time in a series which may not be described well by a single model. The threshold autoregressive (TAR) model \citep{tong,geweke} describes an AR process whose parameters switch according to the value of a previous observation, and is a special case of the Markov switching autoregressive model. 
We refer to \citet{tong1990} for a review of nonlinear time series, and \citet{fruwirth} for a thorough review 
of mixture models for time series. Mixture autoregressive models \citep{juangrabiner,wongli} are also special 
cases of Markov switching AR models, in which the parameters of the autoregression change according to 
a hidden Markov process.

The models discussed above generally achieve nonstationarity or nonlinearity by allowing parameters to switch or evolve in time. These models are naturally suited to problems in which a single parametric model holds in a given interval of time. For instance, the TAR structure assumes only one linear submodel applies at any particular time, with abrupt changes at the thresholds. In contrast, mixture models can be obtained by introducing hierarchical priors on model parameters, to yield a set of parametric models which are favored with different probabilities across time. These models possess the ability to capture features which could not be accommodated under the assumption of a single parametric distribution at a particular point in time. To this end, a mixture modeling approach involving Bayesian nonparametric techniques was first proposed by \citet{mullerwestmac}. More recently, \citet{dilucca} have utilized dependent Dirichlet process priors to build countable mixtures of AR models as well as variations of this model. 
\citet{antoniano} developed stationary time series models which contain general transition and invariant densities. 
Existing mixture models for time series are discussed further in Section \ref{sec:background}, relative to our 
proposed model.

%
Here, we present a general framework for modeling univariate time series data, which 
{assumes time-homogeneity} but makes no assumptions of stationarity, 
and can accommodate complex, nonlinear dynamics as well as non-standard distributions. 
{The proposed model for the transition density takes the form of a 
location-scale mixture of normal densities, with means and mixture weights which 
depend on the previous state(s). This structure arises from the conditional distribution 
implied by a Bayesian nonparametric mixture of bivariate normals.}
Key to the posterior simulation method is a square-root-free Cholesky decomposition of 
the mixture kernel covariance matrix. As demonstrated with synthetic and real data, the 
model enables general inference for time-homogeneous, nonstationary Markovian 
processes indexed in discrete time.

The rest of the paper is organized as follows. The methodology is presented in Section \ref{sec:methods},
including the model formulation for the transition density, and methods for prior specification and 
posterior simulation 
{(technical details for the latter are included in the appendices)}. 
To place our contribution within the relevant literature, we also discuss certain classes of 
mixture models for discrete-time Markovian processes. In Section \ref{sec:applications}, the modeling 
approach is illustrated with simulated data examples, and it is also applied to a standard data set 
on waiting times between successive eruptions of the Old Faithful geyser. 
{For the real data example, we also consider comparison with a parametric 
TAR model and with a more structured version of the proposed mixture model which 
ensures existence of a stationary distribution for the Markov chain; this latter model is 
essentially the one developed by \citet{antoniano}.} 
While the model development and data illustrations are focused on univariate time series 
with first-order dependence, in Section \ref{sec:extensions}, we discuss possible extensions 
to accommodate higher order structure and to develop a state-space model. Finally, 
Section \ref{sec:conclusion} concludes with a summary.

\section{Methodology}
\label{sec:methods}

\subsection{Model Formulation}
\label{sec:model}

Here, we present the model for nonstationary time series. We focus on the case with first-order
Markovian dependence, discussing the extension to modeling higher order time series in 
Section \ref{sec:extensions}. Hence, the observed time series, $(z_{1},\dots,z_{n})$, is assumed to 
be a realization from a time-homogeneous, real-valued, first-order Markov chain, and thus the 
likelihood, conditional on $z_{1}$, is given by $\prod_{t=2}^{n} f(z_{t} \mid z_{t-1})$,
where $f(z_{t} \mid z_{t-1})$ is the transition density.

{{To flexibly model the transition density, we use the conditional density 
$f(y\mid x)$ induced by a nonparametric mixture of bivariate normal distributions for $f(x,y)$.
More specifically, $f(x,y) \equiv$ $f(x,y\mid G)=$
$\int \text{N}(x,y \mid \bm{\mu},\Sigma) \, \text{d}G(\bm{\mu},\Sigma)$, with a Dirichlet process (DP) 
prior \citep{ferguson} for the random mixing distribution $G$.}
In the ensuing model expressions, we work with a truncated version of $G$ motivated by the 
DP constructive definition \citep{sethuraman}, which is also the approach we follow for 
posterior simulation \citep{ishjames}. Under a truncated DP at level $L$, the joint density 
can be expressed as $f(x,y \mid G) =$ $\sum_{l=1}^L p_l\text{N}(x,y \mid \bm{\mu}_l,\Sigma_l)$.
Here, the $(\bm{\mu}_l,\Sigma_l)$ are independent and identically distributed (i.i.d.) from the DP 
base distribution $G_0$, and the weights $(p_1,\dots,p_L)$ are determined through stick-breaking 
from latent beta$(1,\alpha)$ random variables. 
{{
In particular, $p_1=v_1$, $p_l=$ $v_l \prod_{r=1}^{l-1}(1-v_r)$, for $l=2,\dots,L-1$, and 
$p_L=$ $\prod_{r=1}^{L-1}(1-v_r)$, where $v_1,\dots,v_{L-1}\stackrel{i.i.d.}{\sim}\mathrm{beta}(1,\alpha)$.
The choice of the truncation level $L$ is discussed in Section \ref{sec:applications}.}

Partitioning $\bm{\mu}_l$ and $\Sigma_l$ with superscripts $x$ and $y$, the conditional distribution 
for $f(y\mid x,G)$ implied by $f(x,y \mid G)$ is used as the model for the transition density:
\begin{equation}
\label{eqn:trans_dens}
f(z_t \mid z_{t-1}, G) = \sum_{l=1}^L q_l(z_{t-1})\mathrm{N}\left(z_t \mid \mu_l^y+\Sigma_l^{yx}(\Sigma_l^{xx})^{-1}(z_{t-1}-\mu_l^x),\Sigma_l^{yy}-(\Sigma_l^{yx})^2(\Sigma_l^{xx})^{-1}\right)
\end{equation}
with 
\begin{equation}
\label{eqn:weights}
q_l(z_{t-1}) = p_l\mathrm{N}(z_{t-1} \mid \mu_l^x,\Sigma_l^{xx}) /
\left\{ \sum\nolimits_{m=1}^L p_{m} \mathrm{N}(z_{t-1} \mid \mu_m^x,\Sigma_m^{xx}) \right\}.
\end{equation}
The transition density is therefore a location-scale mixture of normal transition densities, 
with means which depend on the previous state in a linear fashion, and weights which favor 
mixture component $l$ if $z_{t-1}$ is near $\mu_l^x$. 

{
The model structure in (\ref{eqn:trans_dens}) and (\ref{eqn:weights}) defines a flexible
time-homogeneous, nonstationary Markov chain model. It allows for very general 
transition density shapes that can change flexibly across the state space, owing to the 
local adjustment provided by the mixture weights. The model also enables rich nonlinear 
dynamic relationships, which can be explored through, for instance, the conditional 
expectation 
$$
\mathrm{E}(Z_t \mid Z_{t-1}=z_{t-1},G) = 
\sum_{l=1}^{L} q_l(z_{t-1}) \{\mu_l^y+\Sigma_l^{yx}(\Sigma_l^{xx})^{-1}(z_{t-1}-\mu_l^x)\}.
$$
This is a mixture of linear functions, but with state-dependent weights which can thus 
uncover nonlinear dynamics, in addition to non-Gaussian transition densities.}

{As discussed above, the transition density in (\ref{eqn:trans_dens}) arises 
from the well-studied DP mixture of normals model.} Conditional on an initial value $z_1$, 
the likelihood $\prod_{t=2}^n f(z_t \mid z_{t-1},G)$ is a product of conditional densities, 
each being a mixture of normals. 
The mixture weights, given by (\ref{eqn:weights}), contain $\{\mu_l^x\}$ and $\{\Sigma_l^{xx}\}$ 
in the denominator, and each mixture component variance in (\ref{eqn:trans_dens}) contains a complex function 
of the elements of $\Sigma_l$. Hence, with respect to posterior simulation, there does not exist a choice 
of $G_0$ which allows the full conditional distributions for $\mu_l^x$, $\Sigma_l^{xx}$, 
$\Sigma_l^{yy}$, or $\Sigma_l^{yx}$ to be recognizable as standard distributions.

These difficulties are alleviated to some extent by employing a square-root-free Cholesky decomposition 
of the covariance matrix $\Sigma$ \citep{daniels,webb,deyoreokottas},
which expresses $\Sigma$ in terms of a unit lower triangular matrix $\beta$ 
and a  diagonal matrix $\Delta$ with positive elements, such that $\Sigma=\beta^{-1}\Delta(\beta^{-1})^T$. 
The utility of this parametrization lies in the following property. If 
$(Y_1,\dots,Y_m)\sim \mathrm{N}(\bm{\mu},\beta^{-1}\Delta(\beta^{-1})^T)$, with 
$(\delta_1,\dots,\delta_m)$ on the diagonal of $\Delta$, then the joint distribution 
of $Y$ can be expressed in a recursive form: $Y_1\sim \mathrm{N}(\mu_1,\delta_1)$, and 
$(Y_k \mid Y_1,\dots,Y_{k-1})\sim \mathrm{N}(\mu_k-\sum_{j=1}^{k-1}\beta_{k,j}(y_j-\mu_j),\delta_k)$, 
for $k=2,\dots,m$. 
With this parameterization of the mixture kernel covariance matrix, the mixture transition density  
(\ref{eqn:trans_dens}) admits the form 
\begin{equation}
\label{eqn:trans_dens_reparam}
f(z_t \mid z_{t-1},G) = \sum_{l=1}^Lq_l(z_{t-1}) \mathrm{N}(z_t \mid \mu_l^y-\beta_l(z_{t-1}-\mu_l^x),\delta_l^y)
\end{equation} 
with 
\begin{equation}
\label{eqn:weights_reparam}
q_l(z_{t-1}) = p_l\mathrm{N}(z_{t-1} \mid \mu_l^x,\delta_l^x) /
\left\{ \sum\nolimits_{m=1}^L p_{m} \mathrm{N}(z_{t-1} \mid \mu_m^x,\delta_{m}^{x}) \right\}
\end{equation}
where, in the case of the $2 \times 2$ covariance matrix $\Sigma$, $\beta$ represents the only 
free element of the lower triangular matrix, and $\Delta$ has diagonal elements $(\delta^x,\delta^y)$.

Let $\bm{\eta}_{l}=$ $(\mu_l^x,\mu_l^y,\beta_l,\delta_l^x,\delta_l^y)$, for $l=1,\dots,L$, denote the mixing parameters.  
The mixture transition density can be broken by introducing latent configuration variables $\{U_2,\dots,U_n\}$ 
taking values in $\{1,\dots,L\}$, with $\mathrm{Pr}(U_t=l)=q_l(z_{t-1})$, such that the augmented hierarchical 
model for the data becomes:
\begin{eqnarray}
\label{eqn:hier_model}
z_t \mid z_{t-1},U_t,\{\bm{\eta}_l\}  \stackrel{ind.}{\sim} 
\mathrm{N}(\mu_{U_t}^y-\beta_{U_t}(z_{t-1}-\mu_{U_t}^x),\delta_{U_t}^y), \,\,\,\,\, t=2,\dots,n \notag \\
U_t \mid z_{t-1},\bm{p},\{\bm{\eta}_l\}  \stackrel{ind.}{\sim} \sum_{l=1}^{L}
\frac{p_l\mathrm{N}(z_{t-1} \mid \mu_l^x,\delta_l^x)}{\sum_{m=1}^Lp_m \mathrm{N}(z_{t-1} \mid \mu_m^x,\delta_m^x)}
I(U_{t} = l),   \,\,\,\,\,  t=2,\dots,n \notag \\
\bm{\eta}_l \mid \bm{\psi} \stackrel{i.i.d.}{\sim} G_0(\bm{\eta}_l \mid \bm{\psi}), \,\,\,\,\,  l=1,\dots,L
\end{eqnarray}
and the prior density for $\bm{p}=$ $(p_1,\dots p_L)$ is given by a special case of the generalized 
Dirichlet distribution: $f(\bm{p} \mid \alpha)=$
$\alpha^{L-1}p_L^{\alpha-1}(1-p_1)^{-1}(1-(p_1+p_2))^{-1}\times\dots\times(1-\sum_{l=1}^{L-2}p_l)^{-1}$
\citep{connor}. The base distribution $G_0$ comprises independent components:  
$\mathrm{N}(m^x,v^x)$ and $\mathrm{N}(m^y,v^y)$ for $\mu_{l}^{x}$ and $\mu_{l}^{y}$;
$\mathrm{IG}(\nu^x,s^x)$ and $\mathrm{IG}(\nu^y,s^y)$ for $\delta_l^x$ and $\delta_l^y$;
and $\mathrm{N}(\theta,c)$ for $\beta_{l}$.
This choice is conjugate for $\{\delta_l^y\}$, $\{\beta_l\}$, and $\{\mu_l^y\}$. The full Bayesian model 
is completed with conditionally conjugate priors on $\bm{\psi}=$ $(m^x,v^x,m^y,v^y,s^x,s^y,\theta,c)$, 
the hyperparameters of $G_{0}$:
\begin{eqnarray}
\label{eqn:priors}
m^x\sim \mathrm{N}(a_m^x,b_m^x), \; m^y\sim \mathrm{N}(a_m^y,b_m^y), \; v^x\sim \mathrm{IG}(a_v^x,b_v^x), \; v^y\sim \mathrm{IG}(a_v^y,b_v^y), \notag \\
s^x\sim \mathrm{Ga}(a_s^x,b_s^x), \; s^y\sim\mathrm{Ga}(a_s^y,b_s^y), \; \theta\sim \mathrm{N}(a_\theta,b_\theta), \; c\sim\mathrm{IG}(a_c,b_c)
\end{eqnarray}
and a gamma prior for the DP precision parameter, $\alpha\sim \mathrm{Ga}(a_{\alpha},b_{\alpha})$.

\subsection{Posterior Inference}
\label{sec:MCMC}

Samples from the posterior distribution 
of model (\ref{eqn:hier_model}) are obtained using a combination of Gibbs sampling 
and Metropolis-Hastings steps. 
{
Details of the Markov chain Monte Carlo (MCMC) algorithm are provided in Appendix A,}
focusing particular attention on vector $(p_1,\dots,p_L)$, which requires the most care in 
developing an effective sampling strategy.

Using the posterior samples, full inference is readily available for 
the transition density, $f(z_{t+1} \mid z_t,G)$, for any value of $z_{t}$. In particular, point and 
interval estimates can be obtained for the forecast distribution, $f(z_{n+1} \mid z_n,G)=$ 
$\sum_{l=1}^{L}q_l(z_n)\text{N}(z_{n+1} \mid \mu_l^y-\beta_l(z_n-\mu_l^x),\delta^{y}_l)$.
The posterior mean estimate corresponds to the posterior predictive 
density for the next observation, since it can be shown that 
$p(z_{n+1} \mid \text{data})=$ $\mathrm{E}\{ f(z_{n+1} \mid z_n,G) \mid \text{data} \}$. 
Point estimates for forecasts further than one step ahead may be obtained fairly easily, and entire 
distributions are also available, albeit at somewhat greater computational expense.

{
It may also be of interest to compare the predictive performance of the model with 
alternative models through one-step-ahead predictive distributions, 
$p(z_{t} \mid \bm{z}_{(t-1)})$. Here, $\bm{z}_{(m)}$ denotes the observed series up to time $m$,
for $m=2,...,n$, such that $\bm{z}_{(n)}$ corresponds to the full data vector.
As detailed in Appendix B, it is possible to compute the value of the posterior predictive 
density $p(z_{t} \mid \bm{z}_{(t-1)})$ at any observed $z_{t}$, using the posterior samples
from fitting the model once to $\bm{z}_{(n)}$. We use these one-step-ahead posterior 
predictive ordinates to supplement graphical model comparison for the data example 
of Section \ref{oldfaithful-data}.}

\subsection{Prior Specification}
\label{sec:priors}

{
We discuss prior specification for the hyperparameters $\bm{\psi}$ of $G_0$, aiming to 
select appropriately diffuse priors which use only a small amount of prior information.
Recall that $\mathrm{E}(Z_t\mid Z_{t-1}=z_{t-1},G)=$
$\sum_{l=1}^Lq_l(z_{t-1})(\mu_l^y+\beta_l\mu_l^x-\beta_lz_{t-1})$. As a default approach,
we assume that on average $Z_{t-1}$ does not inform $Z_t$ in the prior, 
so that $\mathrm{E}(\beta_l)=a_{\theta}=0$. To fix $b_{\theta}$, $a_c$, and $b_c$, the 
parameters contributing to $\mathrm{Var}(\beta_l)$, we note that the $\beta_l$ parameters 
can be thought of as component-specific autoregressive coefficients, and that
stationarity for each Gaussian mixture component requires $|\beta_l|<1$. We thus
select $b_{\theta}$, $a_c$, and $b_c$ such that $\mathrm{Var}(\beta_l)=$
$b_{\theta} + \{ b_c/(a_c-1) \} = 1$ to favor a prior for the $\beta_l$ that places most 
mass on values in the stationary region, but also allows for nonstationarity in the 
mixture components. Here, we set $a_c$ to a small value that ensures finite mean 
for the $\mathrm{IG}(a_c,b_c)$ prior distribution, and we follow a similar approach for 
the shape parameters of the other inverse-gamma priors.}

{
Let $d$ and $r$ be a proxy for the center and range of the data, respectively. 
Based again on the conditional expectation, $\mu_l^y$ can be centered 
around $d$ with variance that is consistent with the scale of the data; for instance,
$a_m^y=d$ and $\mathrm{Var}(\mu_l^y)=$ $b_m^y + \{ b_{v}^{y}/(a_v^y-1) \}=$ $(r/4)^2$. 
Similarly, $\delta_l^y$ can be centered at a value representing the data variance,
that is, $\mathrm{E}(\delta_l^y)=$ $a_s^y \{ b_s^y(\nu^y-1) \}^{-1}=$ $(r/4)^2$. With 
$a_v^y$, $\nu^y$, and $a_s^y$ fixed at relatively small values, $b_m^y$, $b_v^y$, and 
$b_s^y$ can be specified from these expressions.}
 
{
Finally, the $\mu_l^x$ and $\delta_l^x$ parameters correspond to the means and variances of 
the Gaussian densities that define the mixture weights in (\ref{eqn:weights_reparam}).
The parameters $\delta_l^x$ control how quickly the weights decay as $z_{t-1}$ gets farther from 
$\mu_l^x$. Component $l$ receives large weight for $z_{t-1}$ around $\mu_l^{x}$, with the relative 
weight decreasing according to a Gaussian distribution. Hence, using again the rough values for the 
center and range of the data, we set $\mathrm{E}(\mu_l^x)=a_m^x=d$, 
$\mathrm{Var}(\mu_l^x)=$ $b_m^x + \{ b_{v}^{x} / (a_v^x-1) \}=$ $(r/4)^2$, and 
$\mathrm{E}(\delta_l^x)=$ $a_s^x \{ b_s^x(\nu^x-1) \}^{-1}=$ $(r/4)^2$.}

\subsection{Related Mixture Models for Time Series}
\label{sec:background}

\citet{carvalho,carvalho2006} model nonlinear time series through finite mixtures of 
generalized linear models, or experts, resulting in time series models with transition densities 
similar to (\ref{eqn:trans_dens}). However, they approach the problem from a maximum 
likelihood perspective, and require the use of model selection criteria to determine the 
optimal size of the mixture. \citet{wood} consider parametric mixture modeling for time 
series in which the weights are time-dependent and the lag is unknown.

While Bayesian nonparametric techniques have become extremely popular in density estimation, regression, 
and other applications, they have been used to a lesser extent in the context of time series. \citet{mullerwestmac} 
first made use of DP priors to build a model for nonstationary time series. They propose a finite mixture of 
AR models with local weights, where the parameters of the autoregressions and the parameters of the mixture 
weights arise from a random distribution which is assigned a DP prior. 
%
%
\citet{tang:planning} establish posterior consistency for transition densities which can be expressed 
as DP mixtures of Gaussian AR kernels. \citet{tang:cs} consider a particular version of this class of models, 
involving a hyperbolic tangent transformation of lagged terms.
\citet{dilucca} apply a dependent DP (DDP) mixture \citep{maceachern} for the transition densities, 
focusing mainly on the common weights version of the DDP. The DP atoms arise from a normal distribution 
with means linear on the previous observation. Their primary model is a location mixture 
of AR models, with mixing on the AR parameters.  \citet{mena} construct structured transition 
densities to obtain strongly stationary AR models. \citet{lauso} also considered DP mixtures of AR processes. 
\citet{caron} and \citet{fox} assume DP mixture errors within a DLM framework.

%
%

{
The proposed mixture model can be modified such that the Markov chain has a stationary distribution.
In particular, consider the restricted version of the bivariate normal kernel for the joint DP mixture 
from which the transition density is defined, such that $\mu^x=$ $\mu^y \equiv$ $\mu$ and 
$\Sigma^{xx}=$ $\Sigma^{yy} \equiv$ $\sigma^{2}$. Then, it can be shown that the density 
$f(\cdot \mid G)=$ $\sum_{l=1}^{L} p_{l} \text{N}(\cdot \mid \mu_{l},\sigma^{2}_{l})$ satisfies 
$\int_{A} f(u \mid G) \text{d}u =$ $\int \{ \int_{A} f(y \mid x,G) \text{d}y \} f(x \mid G) \text{d}x$, 
for all measurable $A \subset \mathbb{R}$, that is, $f(\cdot \mid G)$ is a stationary (invariant) 
density. This constraint yields transition density 
\begin{equation}
\label{stationary-mixture}
f(z_t \mid z_{t-1},G) =
\sum_{l=1}^Lq_l(z_{t-1})\mathrm{N}(z_t \mid \mu_l-\beta_l(z_{t-1}-\mu_l),\sigma^2_l(1-\beta_l^2))
\end{equation}
with $q_l(z_{t-1}) \propto p_{l} \mathrm{N}(z_{t-1} \mid \mu_l,\sigma^2_l)$, and $\beta_l\in (-1,1)$. 
This is essentially the model studied by \citet{antoniano}, although the version they implemented 
did not involve mixing over the scale parameter $\sigma$.
The modeling framework of \citet{antoniano} begins much like ours, in that a transition mechanism 
is obtained as the conditional density from a bivariate mixture distribution. The authors do not apply a 
truncation approximation to the mixing distribution, and instead develop a posterior simulation method 
based on introduction of multiple sets of latent variables and a trans-dimensional MCMC algorithm. 
The model developed by \citet{antoniano} was previously proposed by \citet{martinez}, however 
it was then thought to be intractable due to the infinite sum appearing in the denominator of the 
transition density mixture weights.}
Although we utilize a truncation approximation to the DP, the sum in the denominator of the 
weights in (\ref{eqn:weights}) still presents challenges in terms of posterior simulation. We develop
a tractable MCMC algorithm by reparameterizing the covariance matrices in $f(x,y\mid G)$ and
working with the stick-breaking weights to develop a slice sampler which indirectly provides 
samples for $\bm{p}$ (see Appendix A).

{
We refer to the special case of the nonparametric mixture model discussed above as the 
``stationary'' mixture model. However, it is important to note that the particular restriction 
ensures existence of a stationary distribution, but not its uniqueness, which would be required 
to develop conditions for additional properties of the stationary Markov chain, such as ergodicity. 
Some results in this direction are studied in \citet{carvalho} under the mixture of experts 
formulation for the transition density.}

\section{Data Illustrations}
\label{sec:applications}

We now illustrate the proposed model on two simulated data sets (Section \ref{simulated-data})
and apply it to the waiting times between eruptions of the Old Faithful geyser
(Section \ref{oldfaithful-data}). 
{For the real data example, we also consider comparison with a parametric 
TAR model and with the stationary mixture model discussed in Section \ref{sec:background} 
as a special case of the proposed model.}

In all cases, MCMC inference for the nonparametric mixture model was implemented in R, 
saving every $20$-th iteration after burn-in, and with a posterior sample of $5,000$ 
used for inference. 
{
For the Old Faithful data example, for which $n=272$, it took about 2.5 hours to collect 
$50,000$ posterior samples (without particular emphasis on optimizing the MCMC code).}
We follow the approach to prior specification described in Section \ref{sec:priors}.

{The DP truncation level $L$ is specified using the expectation of the partial 
sum of the original DP weights, $\mathrm{E}(\sum\nolimits_{l=1}^{L} p_{l} \mid \alpha)=$
$1 - \{ \alpha/(\alpha + 1) \}^{L}$. This expression can be averaged over the prior for 
$\alpha$ to estimate the marginal prior expectation $\mathrm{E}(\sum\nolimits_{l=1}^{L} p_{l})$, 
which is used to specify $L$ given any desired tolerance level for the approximation.
For instance, under a gamma$(0.5,0.5)$ prior on $\alpha$, 
$\mathrm{E}(\sum_{l=1}^{L} p_l)$ is $0.9997$ with $L=30$ and $0.99999$ with $L=50$. We 
used a value of $L$ in this range for all data examples, and monitored the number of effective 
components to ensure it never reached the upper bound.}

\subsection{Simulated Data}
\label{simulated-data}

{We first consider a data set generated from standard Brownian motion to test the model 
in a nonstationary setting, albeit with linear Gaussian transition densities (Section \ref{Brownian-motion}).
Next, we demonstrate the capacity of the model to uncover non-linear, non-Gaussian dynamics, using 
synthetic data from skew-normal transition densities with varying skewness and dispersion 
(Section \ref{skew-normal}).}

\begin{figure}[t!]
\centering
\begin{tabular}{cc}
\includegraphics[height=3in,width=3in]{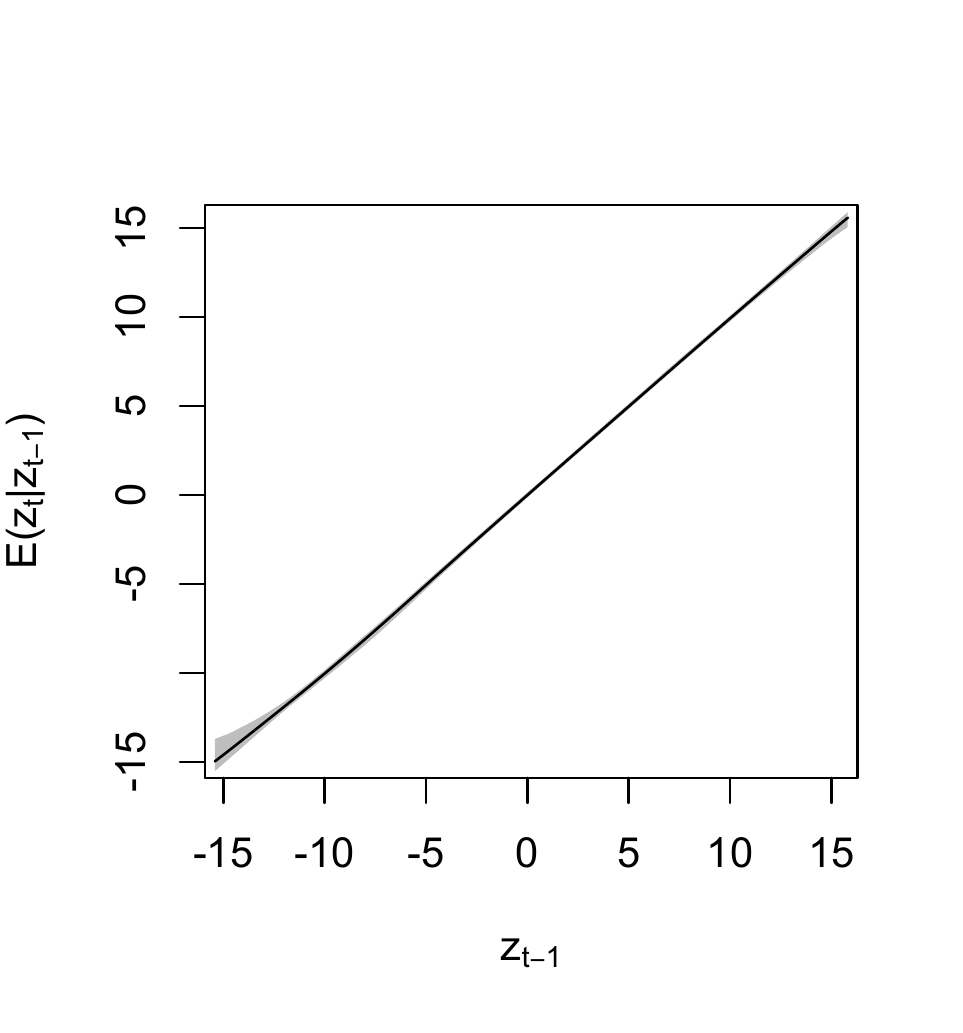}&
\includegraphics[height=3in,width=3in]{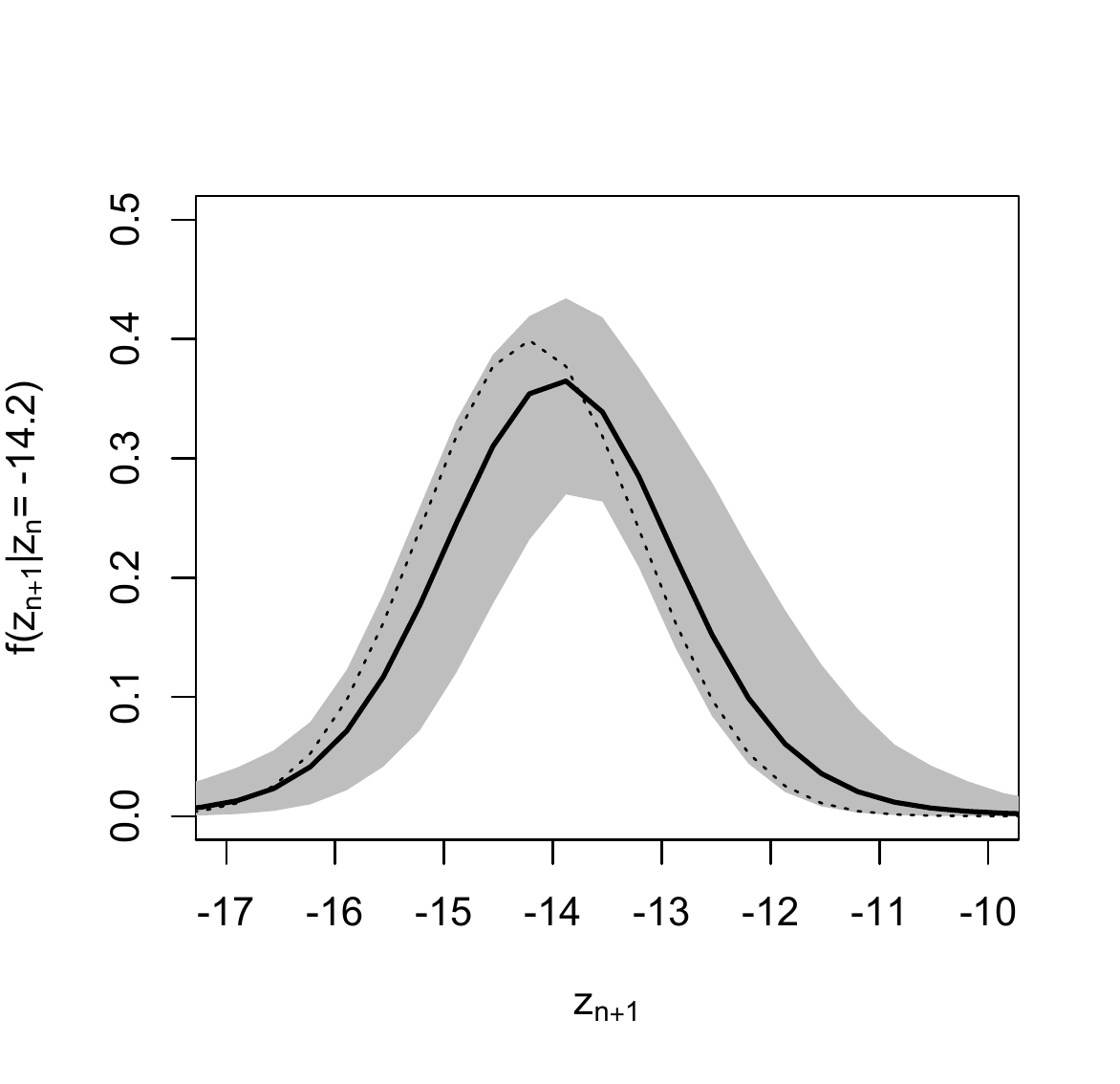}
\end{tabular}
\caption{Brownian motion simulation. The left panel plots the posterior mean estimate (solid line) 
and $95\%$ credible intervals (gray shaded region) for $\mathrm{E}(Z_{t}\mid Z_{t-1}=z_{t-1})$  
plotted over a grid in $z_{t-1}$. The true expectation is indistinguishable from the model's estimate. 
The right panel shows the posterior mean (solid line) and $95\%$ credible intervals (gray shaded 
region) for the forecast density, $f(z_{n+1}\mid z_{n} = -14.2)$, compared to the truth 
(dotted line).}
\label{fig:exp_brownian}
\end{figure}

\subsubsection{Brownian Motion}
\label{Brownian-motion}

Standard Brownian motion is a nonstationary process defined by the transition density $f(z_t\mid z_{t-1})=$
$\mathrm{N}(z_{t-1},1)$. A standard Brownian motion path is generated assuming $n=500$. 
Trivially, $\mathrm{E}(Z_t\mid Z_{t-1}=z_{t-1})=z_{t-1}$ in this model. The inference from the model 
indicates it is detecting this trend with little uncertainty (Figure \ref{fig:exp_brownian}, left panel). 
The value of the last observation is $-14.2$, one of the smallest values in the entire 
series. The forecast density for the next observation is displayed in Figure \ref{fig:exp_brownian} 
(right panel). While the $95\%$ posterior credible intervals contain the true density, the mode of the 
point estimate favors slightly larger values, likely due to the fact that $-14.2$ is an extreme value in 
this series. 

\begin{figure}[t!]
\centering
\includegraphics[height=2.7in,width=5.5in]{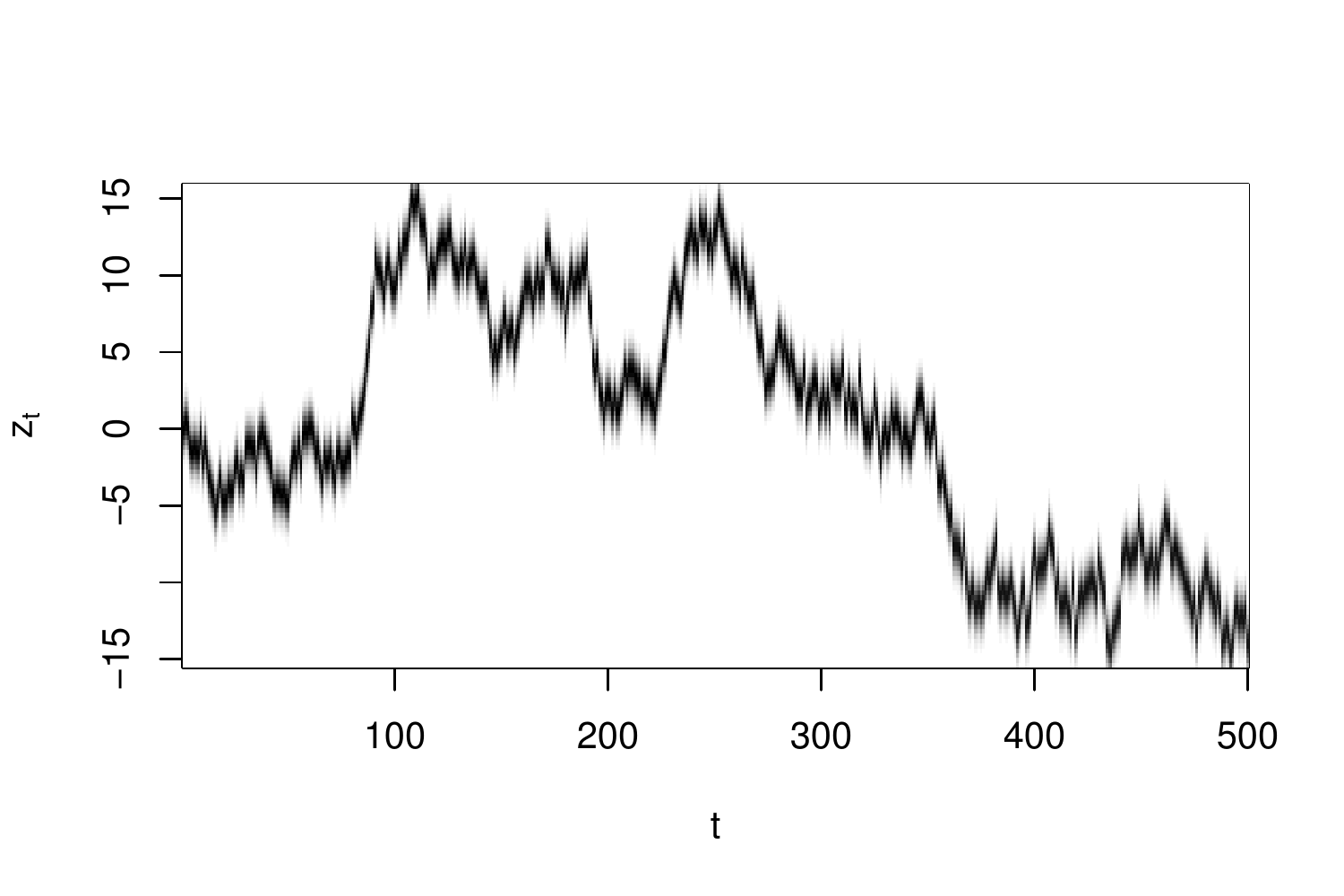}
\includegraphics[height=2.7in,width=5.5in]{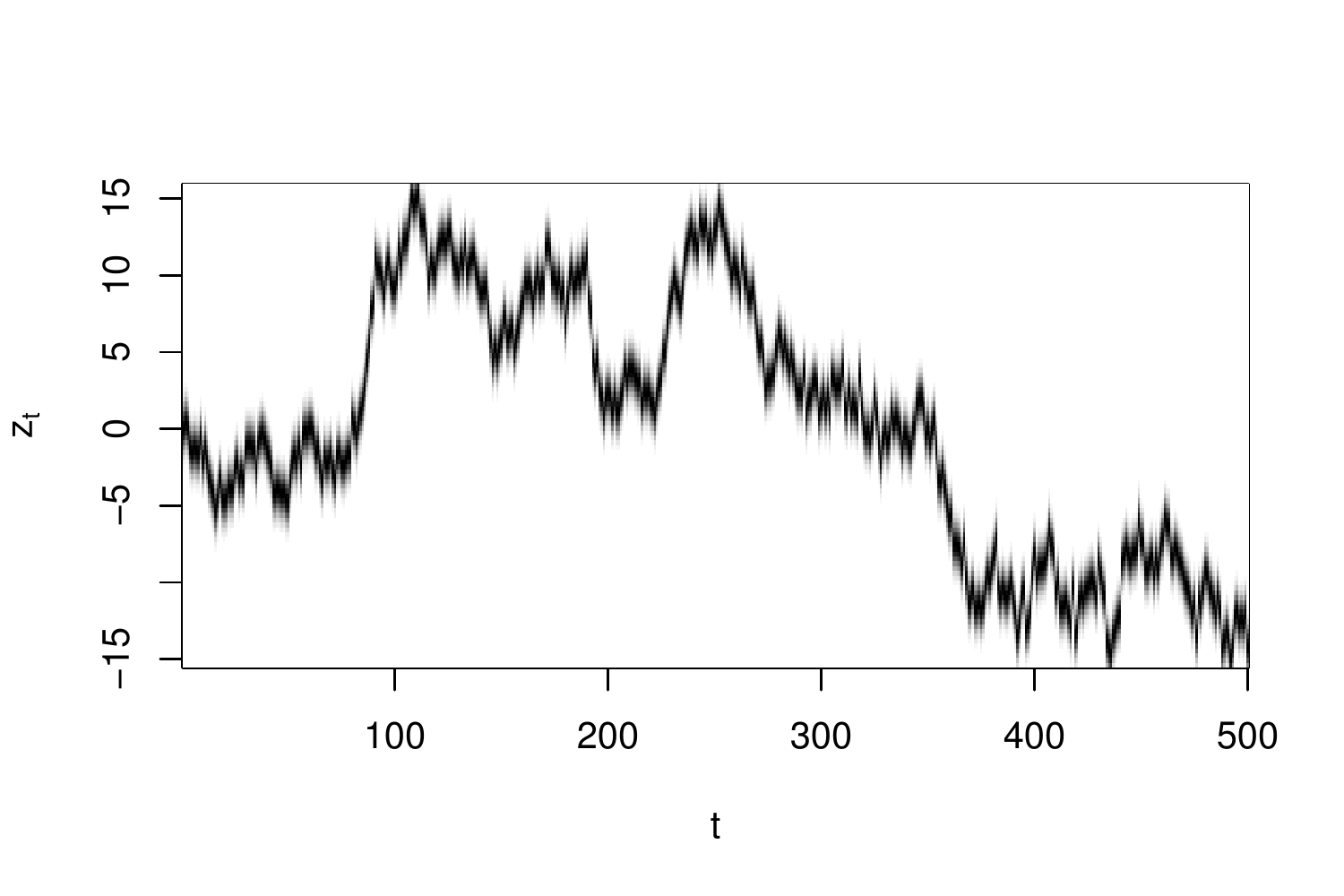}
\caption{Brownian motion simulation. For each $t=2,...,n$, the top panel shows
the posterior mean estimates for $f(z_{t} \mid z_{t-1})$, and the bottom panel plots
the corresponding true densities. Darker colors indicate larger density values.
Refer to Section \ref{Brownian-motion} for details.}
\label{fig:cond_all_brownian}
\end{figure}

{
The top panel of Figure \ref{fig:cond_all_brownian} plots the posterior mean estimates for 
$f(z_{t} \mid z_{t-1})$ for each $t=2,...,n$ and for the corresponding observed $z_{t-1}$.
In particular, for each index $t$ on the horizontal axis, $\text{E}\{ f(z_{t} \mid z_{t-1},G) \mid \text{data} \}$
is plotted on the vertical axis such that darker colors represent larger density values. 
The associated true densities $f(z_{t} \mid z_{t-1})$, again given the observed $z_{t-1}$, are 
plotted in the bottom panel of Figure \ref{fig:cond_all_brownian}.}

{
In summary, all visual displays indicate that the model is capturing the dynamics quite well, 
even though its transition densities are substantially less structured than the transition 
densities of the underlying Brownian motion.}

\begin{figure}[t!]
\centering
\begin{tabular}{cc}
\includegraphics[height=3in,width=2.9in]{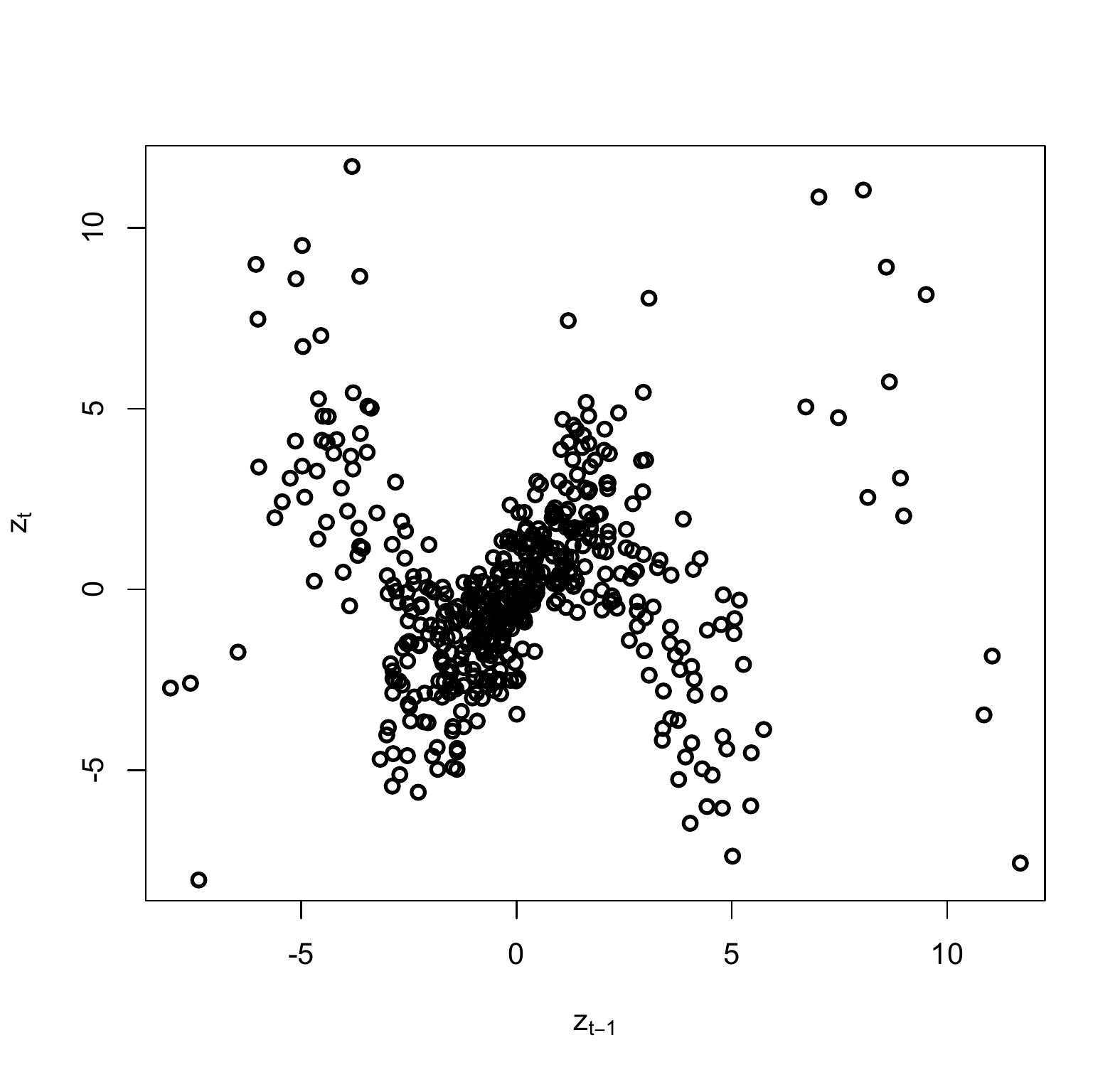}&
\includegraphics[height=3in,width=2.9in]{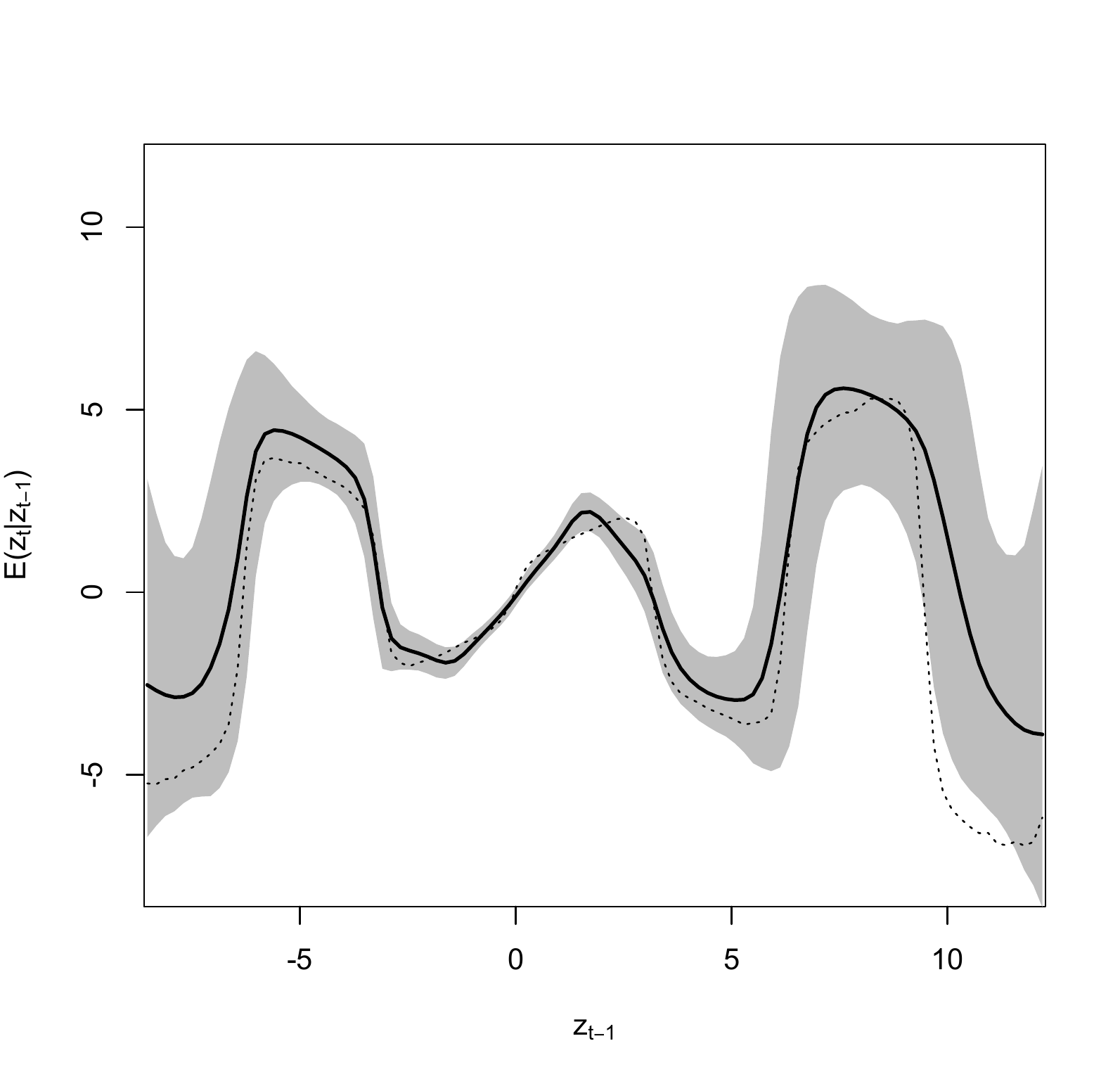}
\end{tabular}
\caption{Skew-normal simulation. The left panel plots the simulated data as pairs of points $(z_{t-1},z_t)$. 
The right panel shows the posterior mean (solid line) and $95\%$ credible intervals (gray shaded region)
for $\mathrm{E}(Z_{t}\mid Z_{t-1}=z_{t-1})$ plotted over a grid in $z_{t-1}$; the true expectation is shown 
as a dotted line.}
\label{fig:SN_lagged}
\end{figure}

\subsubsection{Skew-normal Transition Densities}
\label{skew-normal}

To generate a time series that exhibits challenging transition densities which evolve over time in a 
non-standard fashion, we assume each observation is generated from a skew-normal distribution \citep{azzalini}, 
with scale and skewness parameters which are functions of the previous observation. In particular, 
we generate $z_{t} \mid z_{t-1} \sim$ $\mathrm{SN}(z_t \mid 0,1+0.7|z_{t-1}|,0.1+4\sin(z_{t-1}))$, for $t=2,\dots,n$. 
Here, $\mathrm{SN}(y \mid \xi,\omega,\alpha)$ denotes the skew-normal distribution 
with density $(\omega\pi)^{-1}\exp\{ -(y-\xi)^2/(2\omega^2) \} \Phi(\alpha(x-\xi)/\omega)$, where
$\Phi(\cdot)$ is the standard normal cumulative distribution function. The sinusoidal or periodic 
trend in skewness parameter $\alpha$ yields conditional distributions with various directions and 
degrees of skewness, and the decreasing followed by increasing linear trend in scale parameter $\omega$ 
leads to distributions which are more peaked when $z_{t-1}$ is near $0$.

%
%

A time series $(z_2,\dots,z_{500})$ was simulated from this model assuming an initial value 
$z_1=0$. Figure \ref{fig:SN_lagged} (left panel) shows the simulated
data $\{(z_{t-1},z_t), \, t=2,\dots,500\}$. Notice the oscillating 
trend in location, and the larger variation in $z_t$ for $z_{t-1}$ far from $0$. 
Figure \ref{fig:SN_lagged} (right panel) plots posterior mean and interval estimates for 
$\mathrm{E}(Z_{t}\mid Z_{t-1}=z_{t-1})$ along with the data-generating expectation trend. 
The point estimate captures successfully the overall non-linear trend, and the $95\%$ 
credible intervals contain the truth everywhere except for a small region around 
$z_{t-1}=10$, where there is very little data.

\begin{figure}[t!]
\centering
\begin{tabular}{cc}
\includegraphics[height=2.3in,width=2.5in]{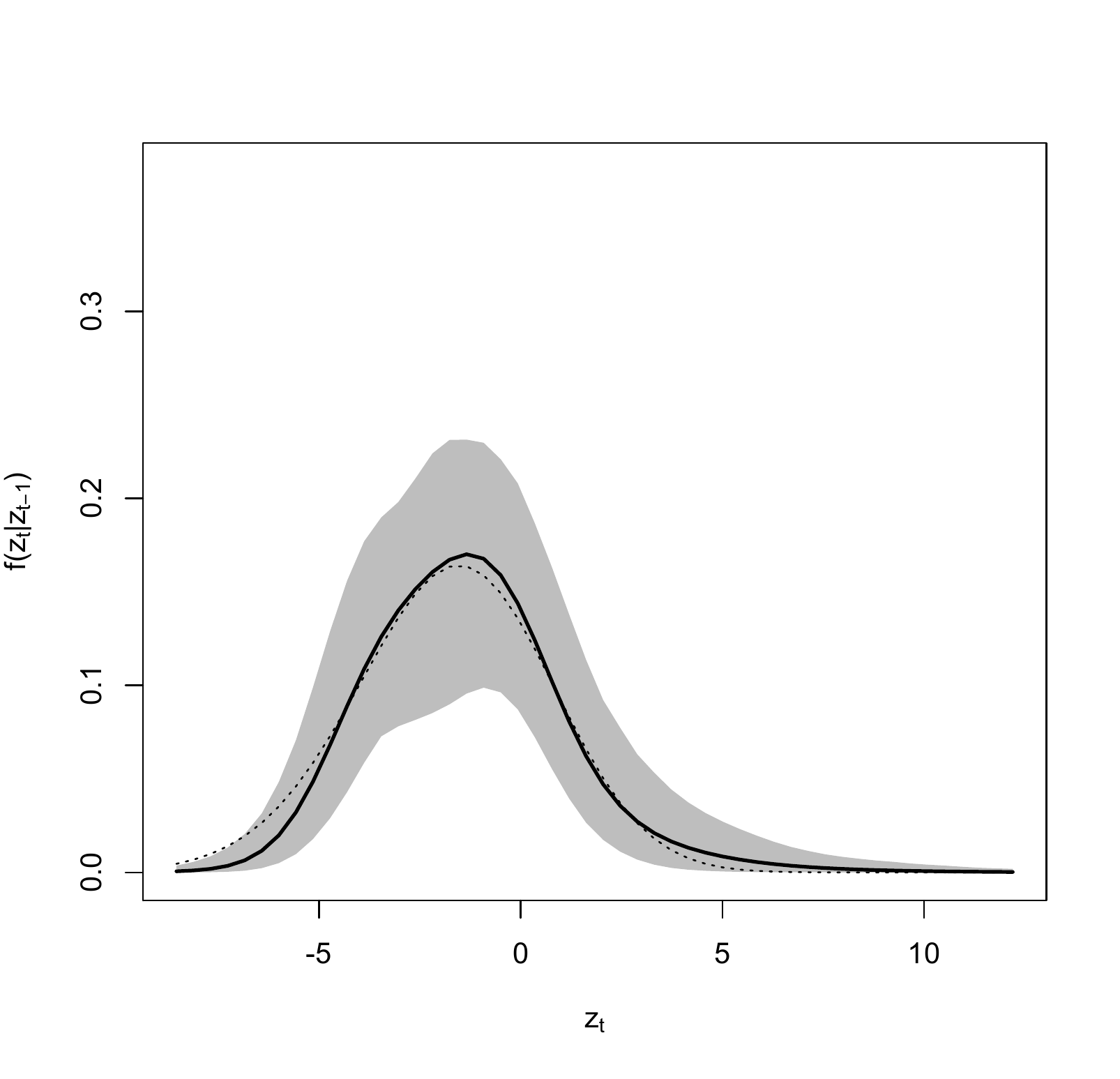}&
\includegraphics[height=2.3in,width=2.5in]{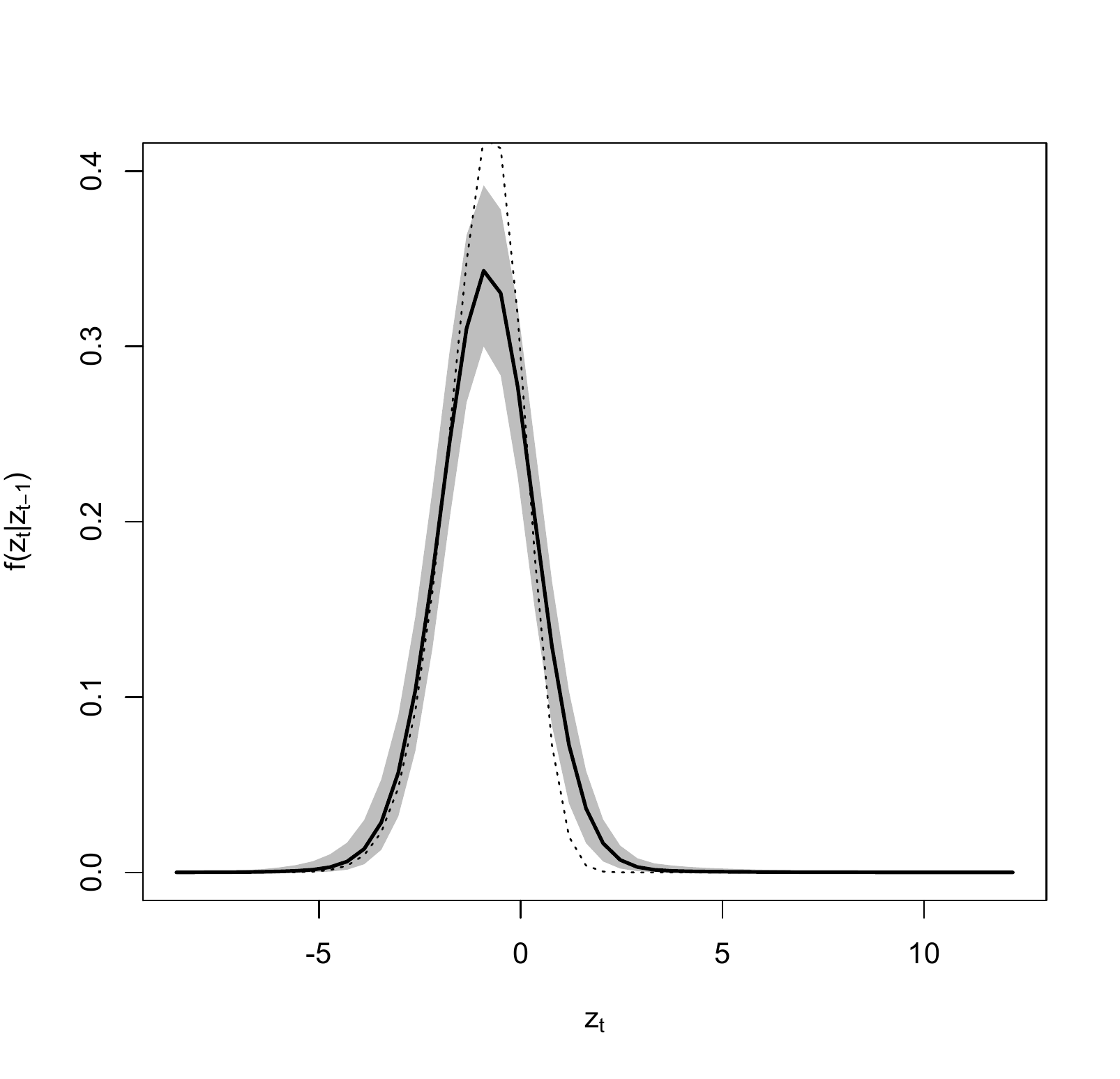}\\
\includegraphics[height=2.3in,width=2.5in]{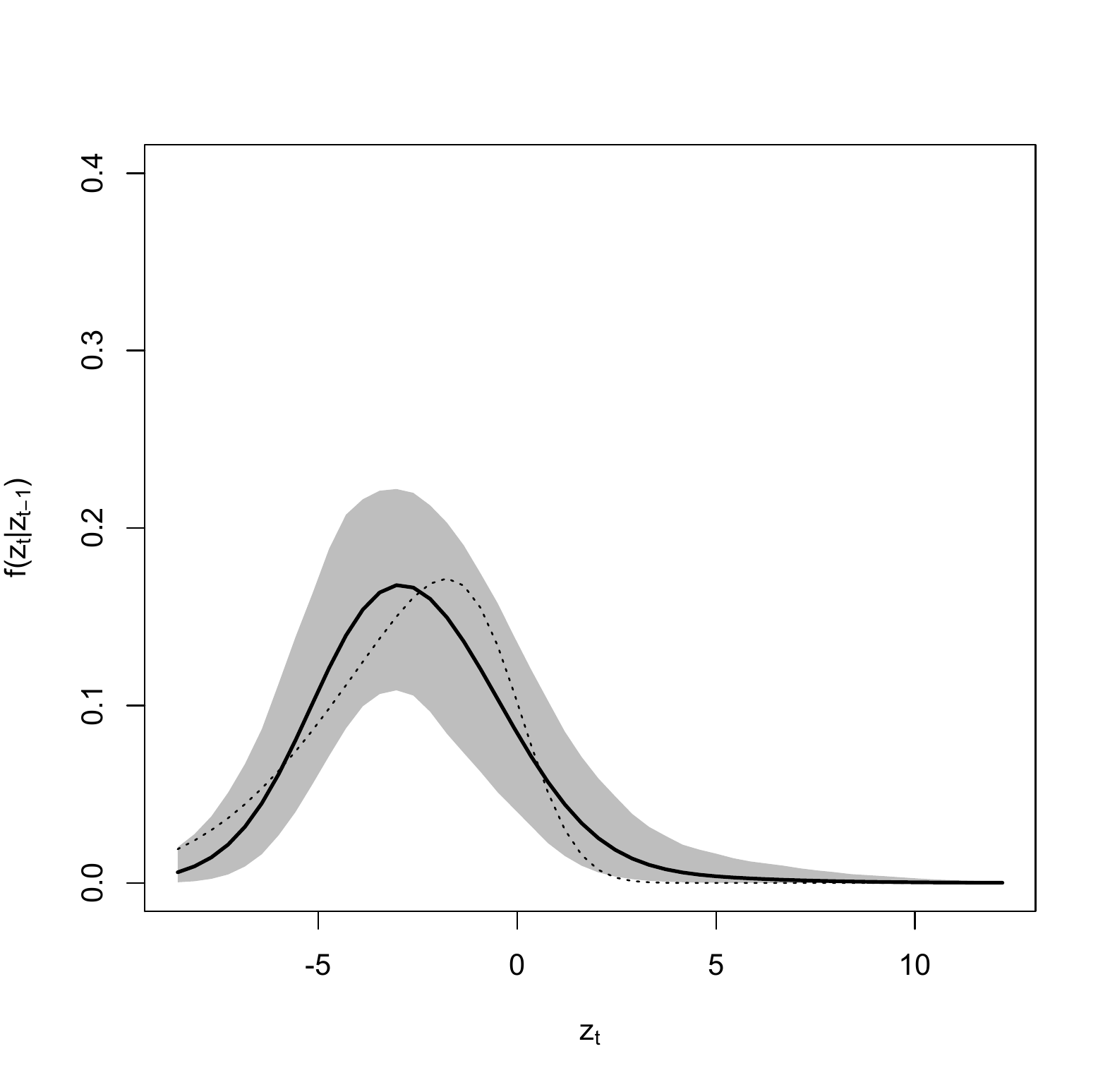}&
\includegraphics[height=2.3in,width=2.5in]{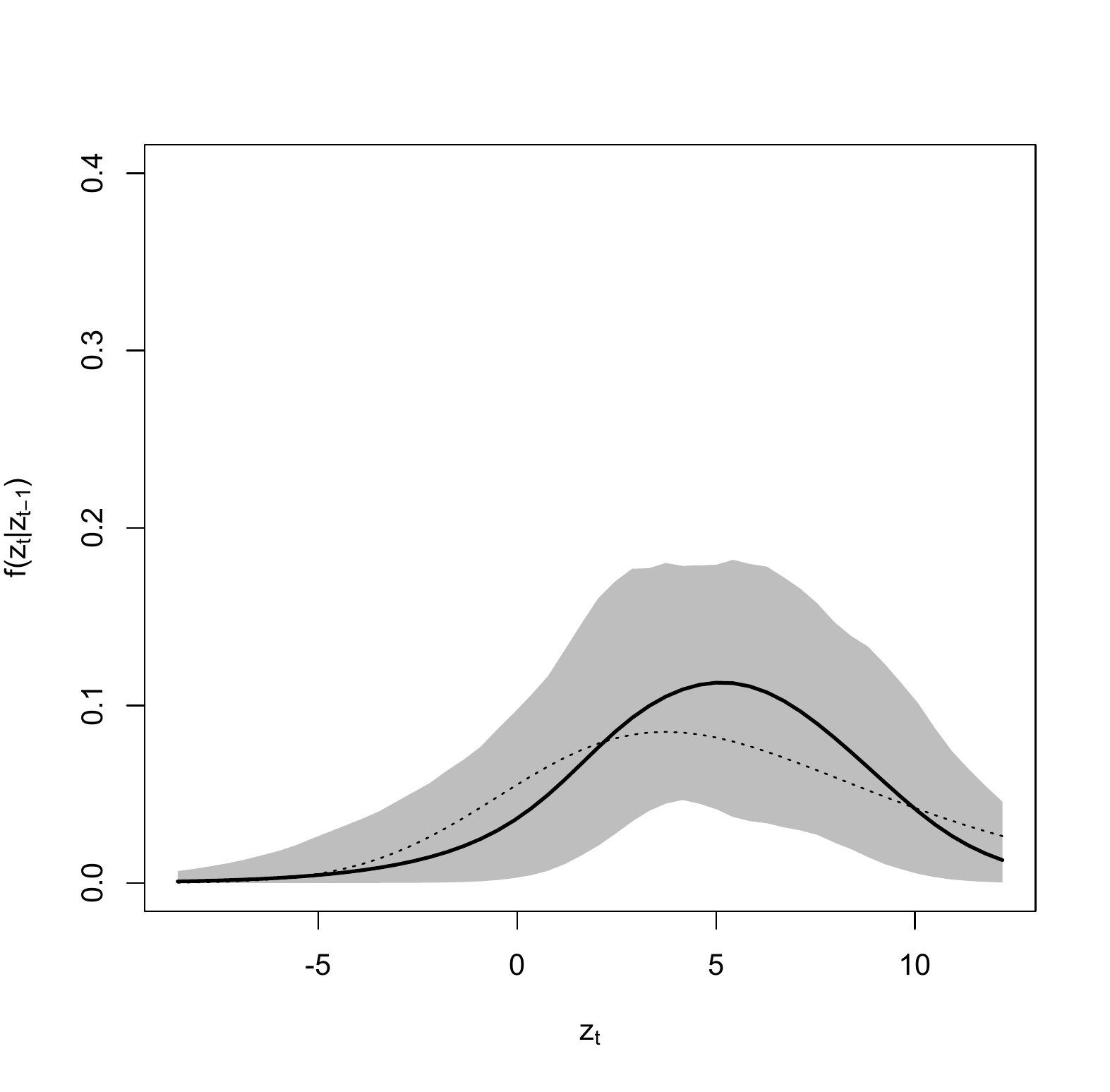}
\end{tabular}
\caption{Skew-normal simulation. Posterior mean (solid line) and $95\%$ credible intervals 
(gray shaded region) for transition densities $f(z_t\mid z_{t-1})$, for $z_{t-1}=-2.85$ (top left), 
$z_{t-1}=-0.5$ (top right), $z_{t-1}=4.2$ (bottom left), and $z_{t-1}=8.85$ (bottom right). 
The corresponding true densities are plotted as dotted lines.}
\label{fig:cond_dens_SN}
\end{figure}

%
%

{
In this case, the true densities $f(z_t\mid z_{t-1})$ do not depict a strong trend analogous to the 
one in Figure \ref{fig:cond_all_brownian}, but the model was again successful in capturing 
the evolution of the skew-normal transition densities through the corresponding posterior 
mean estimates (results now shown). To demonstrate the capacity of the model to 
uncover varying density shapes and the corresponding uncertainty quantification,}
in Figure \ref{fig:cond_dens_SN} we display point estimates and $95\%$ uncertainty 
bands for $f(z_t\mid z_{t-1})$ at four particular values of $z_{t-1}$. 
Notice the wide uncertainty bands for the density at $z_{t-1}=8.85$ (bottom right panel) 
and the narrow uncertainty bands when $z_{t-1}=-0.5$ (top right panel), which reflects 
the lack of data above $z_{t-1}=5$ and the large amount of data in the region near $z_{t-1}=0$.

\subsection{Waiting Times Between Eruptions of the Old Faithful Geyser}
\label{oldfaithful-data}

For our real data illustration, we consider the time intervals between
successive eruptions of the Old Faithful geyser.
The specific data set is available through R (dataset {\tt faithful}) and it consists of $272$ 
measurements $\{z_t, \, t=1,\dots,272\}$, where $z_t$ represents the waiting time 
in minutes before eruption $t$. The data are included in Figure \ref{fig:faithful_exp} in 
the form of a plot of $z_t$ versus $z_{t-1}$, for $t=2,\dots,272$.  

%
%
%

There are some interesting features present in the data. When $z_{t-1}$ is below $60$, there is a large 
cluster of points around $z_t=80$, and a small number of points extending down to about $z_t=50$, 
indicating a distribution with a mode near $80$ but with a heavy left tail or a small additional mode 
near $50$. Moving to larger values of $z_{t-1}$, there are two clusters of points, one centered around 
$55$ and one around $80$. These features are captured by the mixture
model in (\ref{eqn:trans_dens_reparam}) and (\ref{eqn:weights_reparam}), as shown 
in Figure \ref{fig:faithful_cond} with the estimated transition densities $f(z_t \mid z_{t-1}=50)$ and 
$f(z_t \mid z_{t-1}=80)$. 
{
Moreover, inference for $\mathrm{E}(Z_t\mid Z_{t-1}=z_{t-1})$ is shown in 
Figure \ref{fig:faithful_exp} (right panel), and the posterior mean estimate and $95\%$ credible 
intervals for the forecast density, $f(z_{n+1}\mid z_{n}=74)$, are given in Figure 
\ref{fig:faithful_forecast} (right panel).
The estimated forecast density has a primary mode near $80$ and a heavy left tail with a suggestion of 
additional modes around $55$ and $65$; this is a plausible shape for the 
forecast density given the cross-section of data around the last observation $z_{272}=74$.}

\begin{figure}[t!]
\centering
\begin{tabular}{cc}
\includegraphics[height=2.5in,width=2.5in]{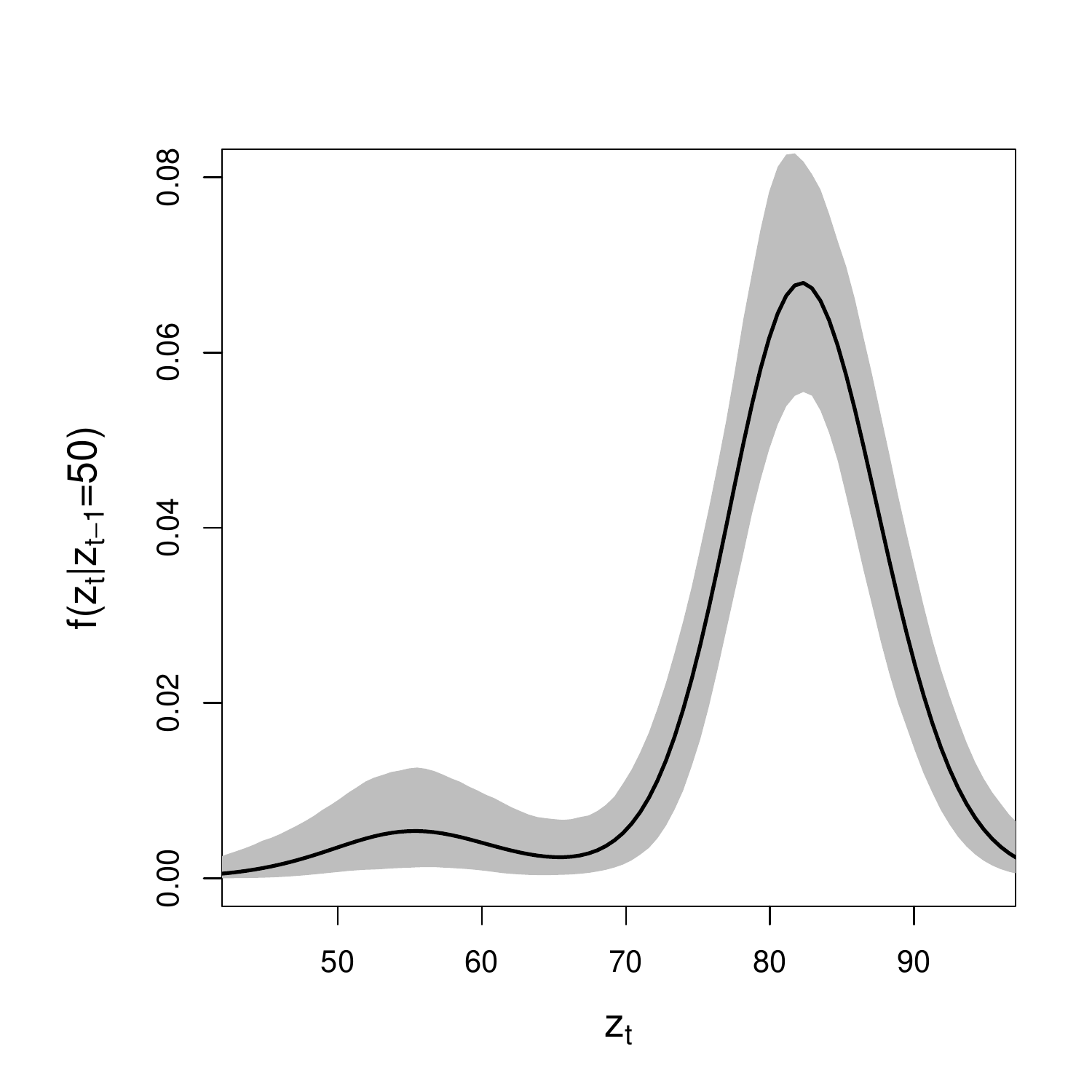}&
\includegraphics[height=2.5in,width=2.5in]{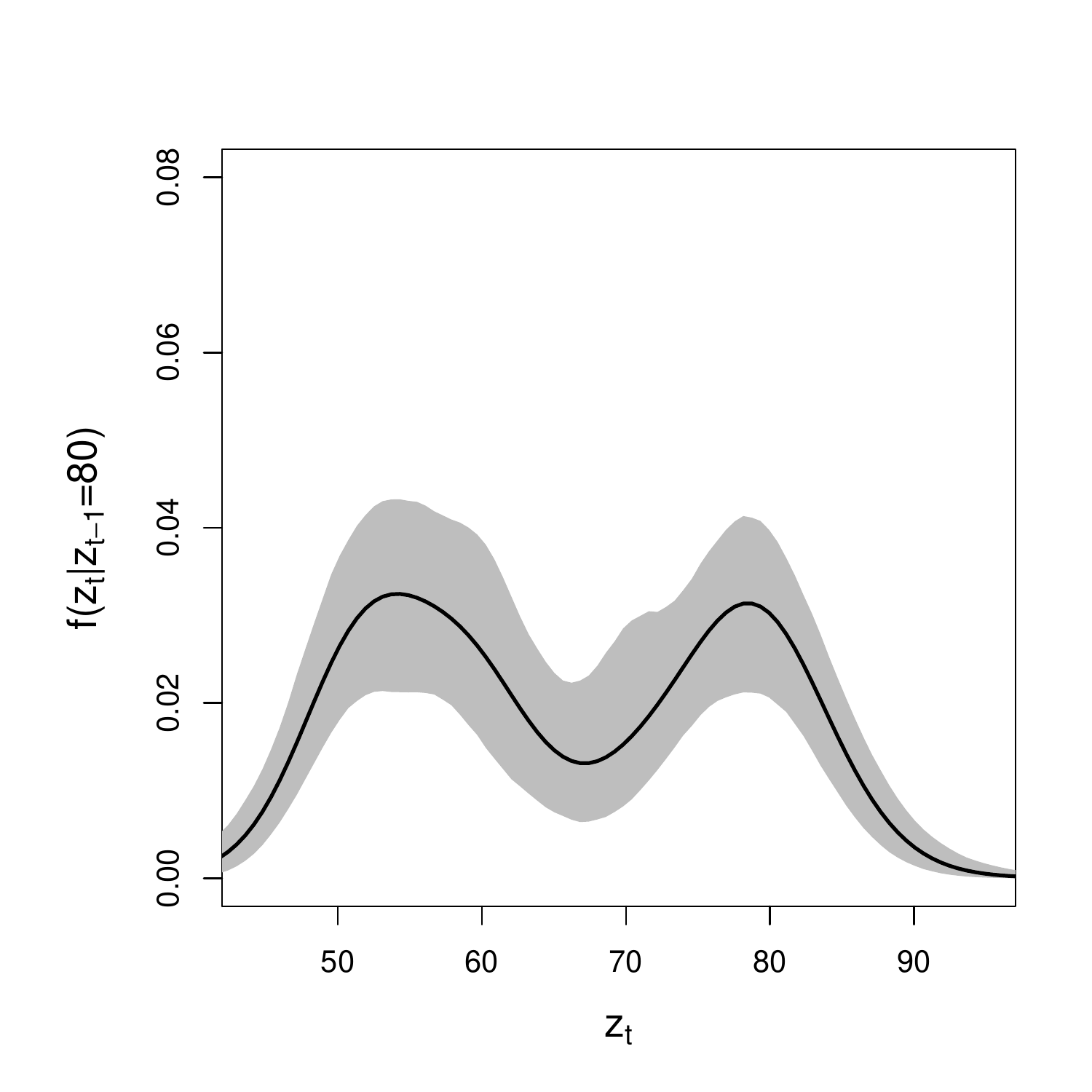} 
\end{tabular}
\caption{Old Faithful data. Posterior mean (solid line) and $95\%$ credible intervals 
(gray shaded region) for transition densities $f(z_t\mid z_{t-1})$, at $z_{t-1}=50$ 
(left panel) and $z_{t-1}=80$ (right panel), under the general mixture model.}
\label{fig:faithful_cond}
\end{figure}

Next, we discuss results from comparison with a parametric model and with the 
special case of the proposed model that incorporates the stationarity restriction.
The stationary mixture model, given in (\ref{stationary-mixture}), was implemented 
by appropriately modifying the MCMC algorithm described in Appendix A, and with 
priors that were comparable to the ones used for the general mixture model.
Regarding the choice of a parametric model for comparison, 
inspection of the data suggests the TAR model structure 
as a plausible, simpler alternative to capture the nonlinear dynamics in 
$\mathrm{E}(Z_t\mid Z_{t-1}=z_{t-1})$. We thus consider a Gaussian-based TAR model 
with threshold that depends on the previous value in the time series, and with two regimes.
More specifically, $z_t \mid z_{t-1} \sim \mathrm{N}(\phi_{0}^{(1)}+\phi_1^{(1)} z_{t-1},\tau^{(1)})$ 
if $z_{t-1} \leq r$, and $z_t \mid z_{t-1} \sim\mathrm{N}(\phi_{0}^{(2)}+\phi_1^{(2)} z_{t-1},\tau^{(2)})$ 
if $z_{t-1}>r$. The model was implemented with the R package BAYSTAR, ``Bayesian analysis 
of threshold autoregressive models''. We used data-based, informative priors, in particular: the 
prior on the intercept parameters was normal centered at the midpoint of the time series, 
with variance equal to the approximate variance of the time series;
the AR coefficient parameters were given normal priors with mean 0 and variance 2; and 
the $\tau$ parameters were assigned inverse-gamma priors centered at the 
residual mean square error of fitting an AR(1) model to the data, and with small shape 
parameters. The posterior means of the AR coefficients were $0.17$ and $-0.68$, and 
the posterior mean for the threshold $r$ was $64.4$.

\begin{figure}[t!]
\centering
\begin{tabular}{ccc}
\includegraphics[height=2.55in,width=2in]{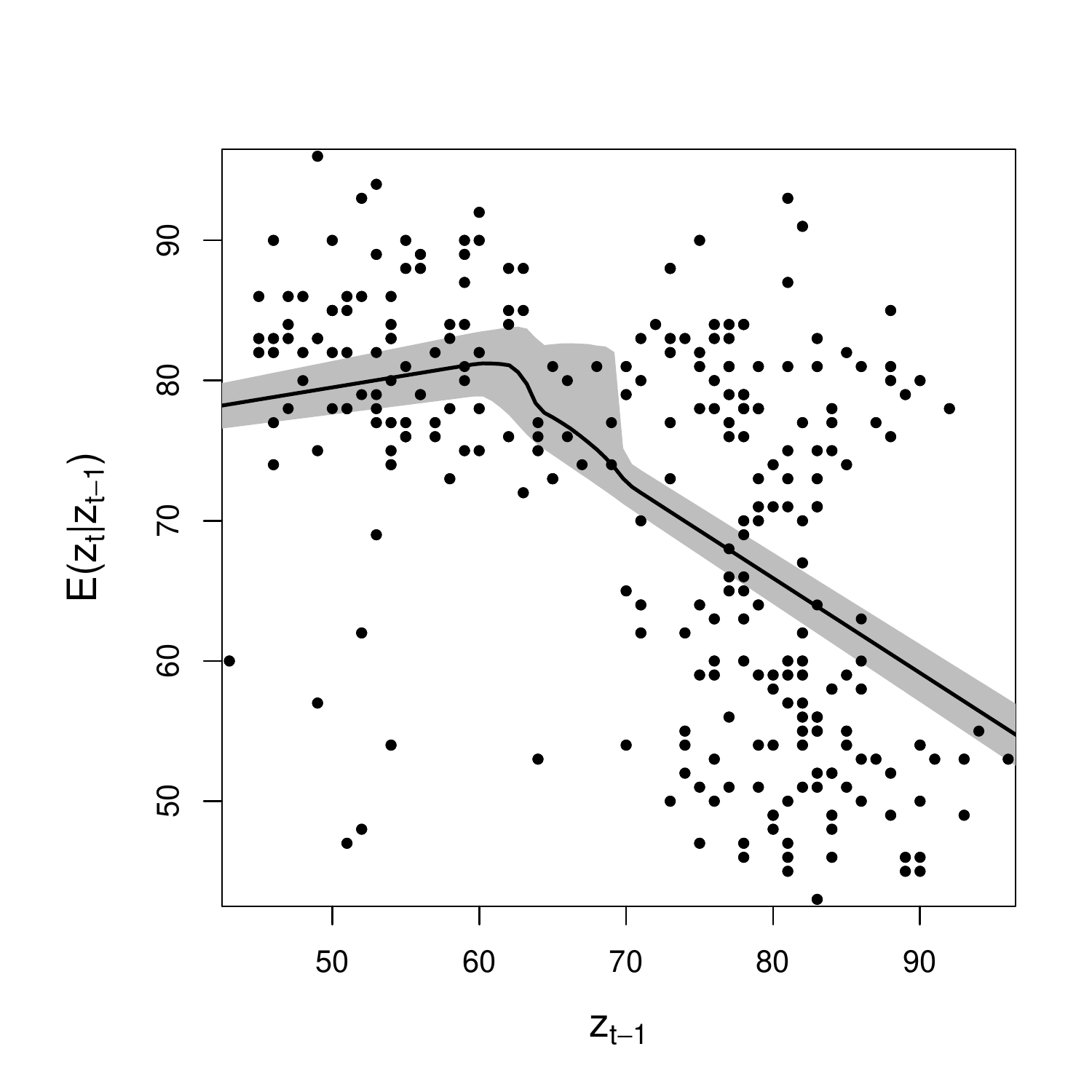}&
\includegraphics[height=2.55in,width=2in]{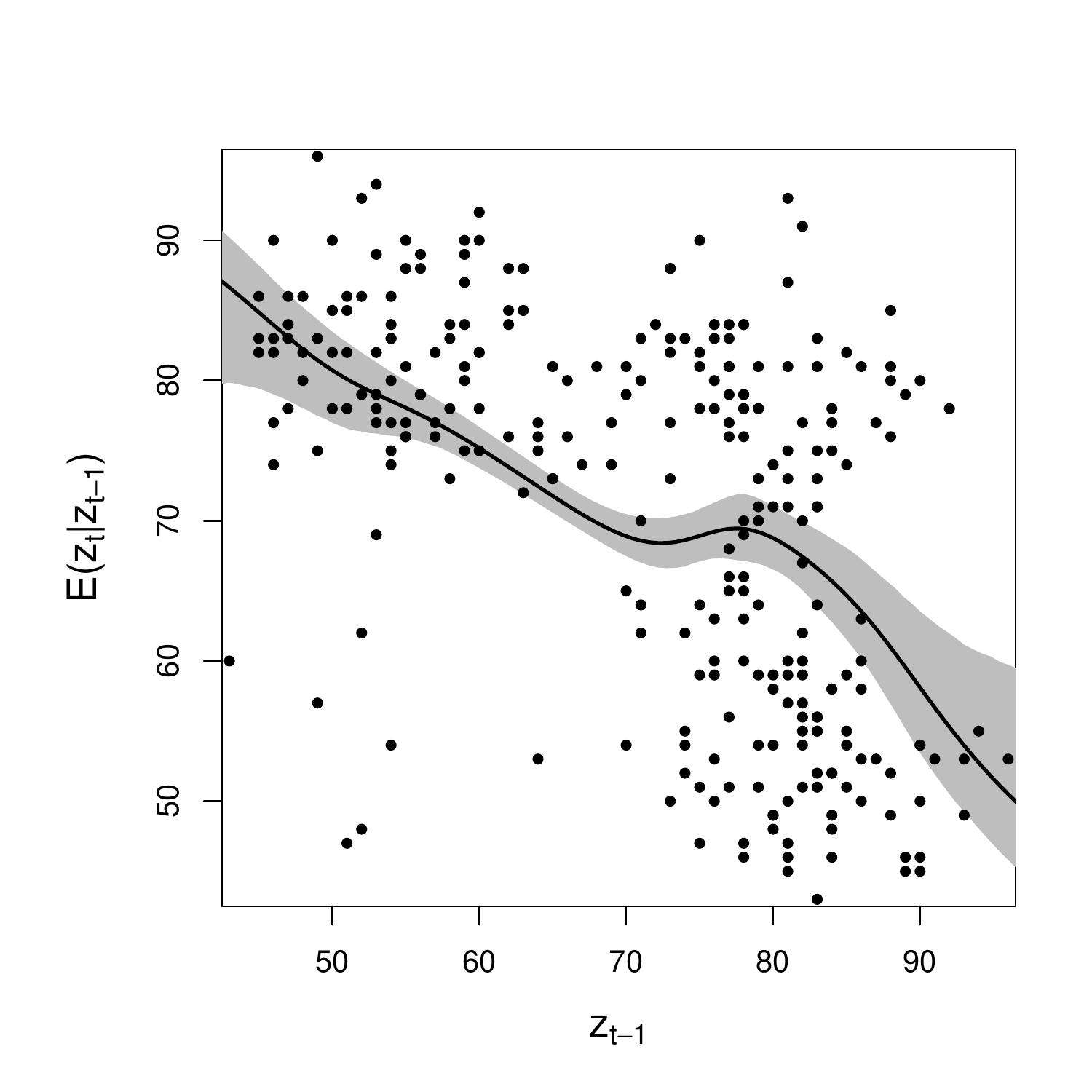}&
\includegraphics[height=2.55in,width=2in]{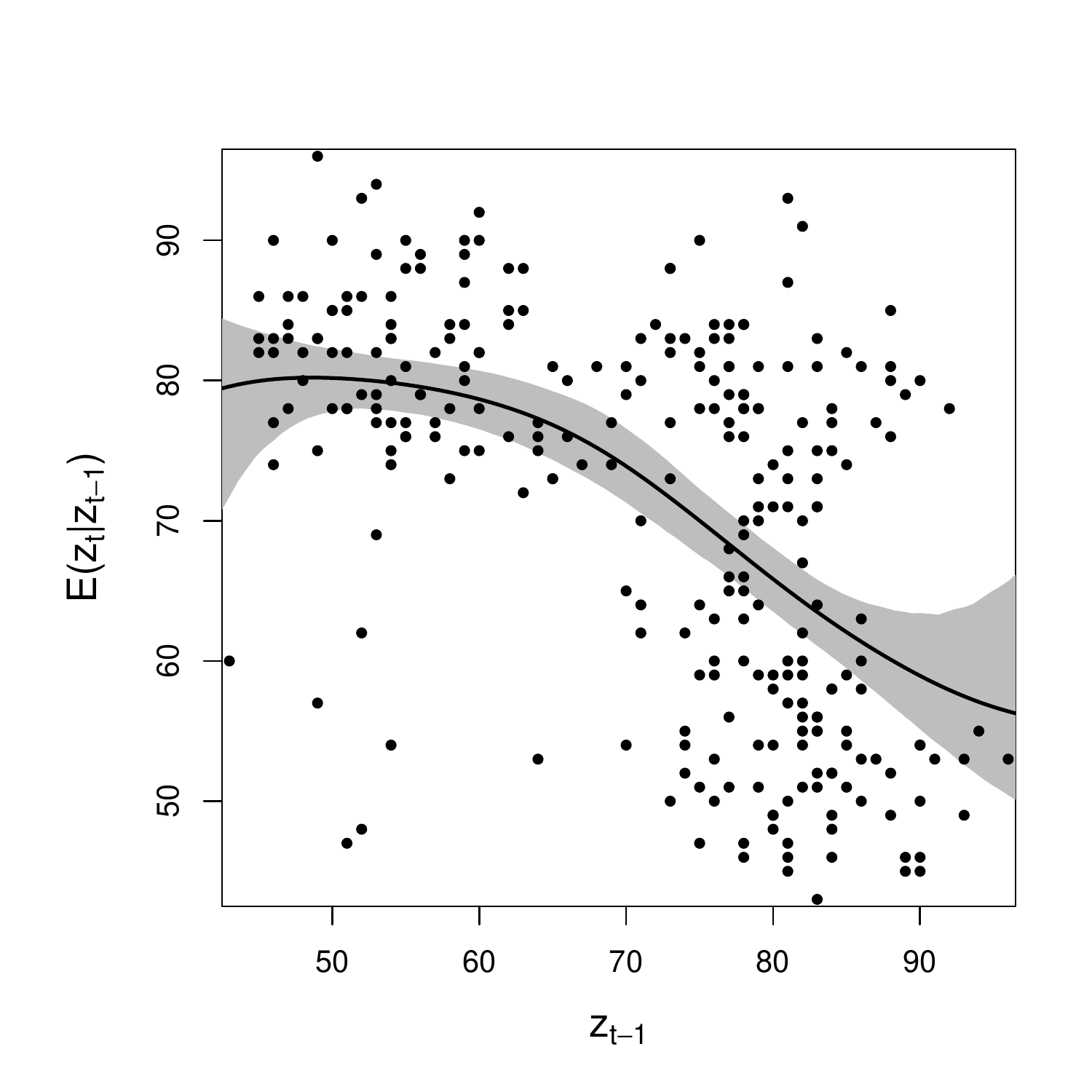}
\end{tabular}
\caption{
Old Faithful data. Posterior mean (solid line) and $95\%$ credible intervals 
(gray shaded region) for $\mathrm{E}(Z_{t}\mid Z_{t-1}=z_{t-1})$ plotted over a grid in $z_{t-1}$,
under the TAR model (left panel), the stationary mixture model (middle panel), and the 
general mixture model (right panel). Included in each panel are the data shown as 
pairs of points $(z_{t-1},z_{t})$, for $t=2,\dots,272$.}
\label{fig:faithful_exp}
\end{figure}

Inference results for $\mathrm{E}(Z_t\mid Z_{t-1}=z_{t-1})$ and for the forecast density 
are reported in Figures \ref{fig:faithful_exp} and \ref{fig:faithful_forecast}, respectively. 
Focusing first on the conditional expectation estimates, the TAR model
uncovers a nonlinear shape that is overall comparable to the one estimated by
the general mixture model, although the latter produces a smoother
point estimate and uncertainty bands that increase at the data
boundaries. The stationary mixture model also yields more plausible 
uncertainty quantification than the parametric model, but estimates a
nonlinearity for $\mathrm{E}(Z_t\mid Z_{t-1}=z_{t-1})$ at a range of $z_{t-1}$
values that is distinctly different from the other two models.
Regarding the forecast density estimates, the TAR model is unable to
capture the non-standard shape uncovered by the general mixture
model. The stationary mixture model estimates a bimodal forecast
density, but with a significant difference in the magnitude of the
peaks relative to the unrestricted mixture model; in particular, the
more pronounced mode around $65$ does not seem to be compatible with
the cross-section of data around $z_{272}=74$. The superior predictive 
performance of the general mixture model is further supported by 
one-step-ahead predictions. Using the approach described in Appendix B, 
we computed the one-step-ahead posterior predictive ordinates for the
last 90 observations (about 1/3 of the observed time series). The sum
of the log-ordinates was $-364.9$ for the stationary mixture model, 
and $-327.4$ for the unrestricted mixture model. The corresponding 
value for the TAR model was $-344.0$, that is, based on this criterion, 
the parametric model performs better than the stationary mixture model.

\begin{figure}[t!]
\centering
\begin{tabular}{ccc}
\includegraphics[height=2.5in,width=2in]{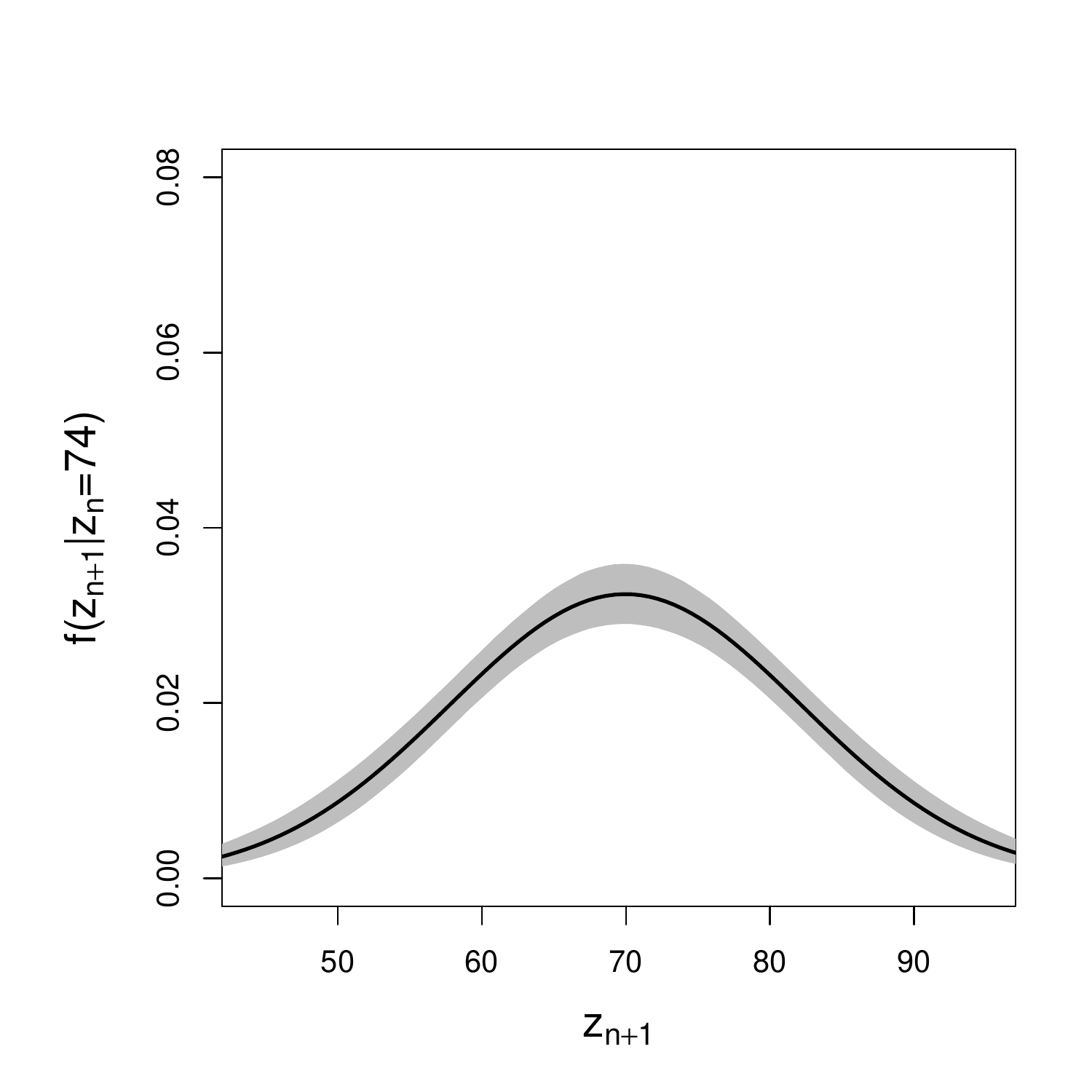}&
\includegraphics[height=2.5in,width=2in]{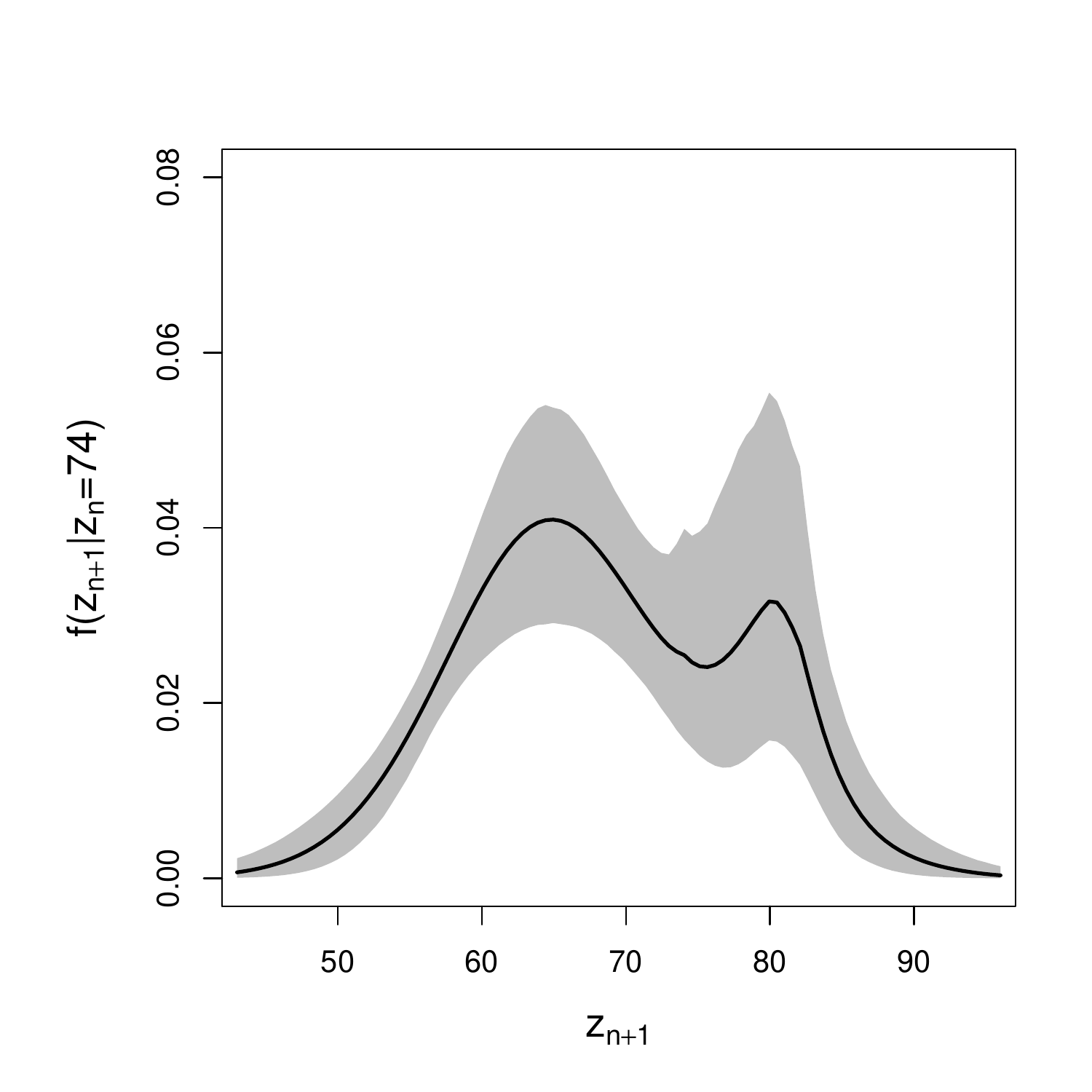}&
\includegraphics[height=2.5in,width=2in]{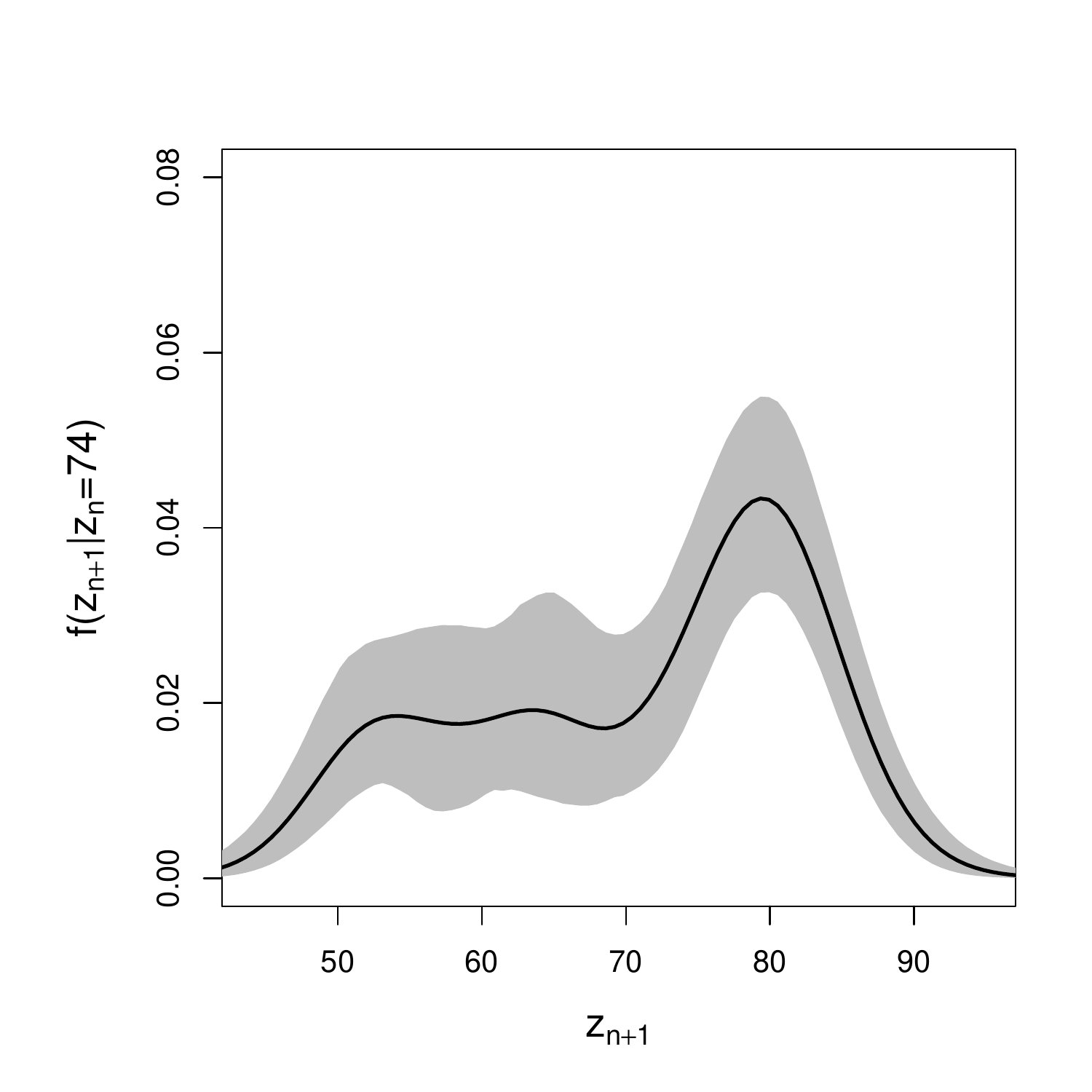}
\end{tabular}
\caption{
Old Faithful data. Posterior mean (solid line) and $95\%$ credible intervals 
(gray shaded region) for the forecast density, $f(z_{n+1}\mid z_{n}=74)$, under the 
TAR model (left panel), the stationary mixture model (middle panel), and the 
general mixture model (right panel).}
\label{fig:faithful_forecast}
\end{figure}

For problems where one has information regarding stationarity of the
data generating process, the stationary mixture model may provide a
natural starting point for the analysis. In fact, as demonstrated in 
\citet{antoniano}, this model is able to estimate effectively transition 
densities from nonstationary processes, which however are driven by
standard distributions. This was confirmed by a reanalysis of the
Brownian motion simulated data example -- which is also one of the
examples of \citet{antoniano} -- for which we obtained from the
stationary mixture model results very similar to the ones reported 
in Section \ref{Brownian-motion}.
However, our experience suggests that the stationary mixture model 
may not be sufficiently flexible for settings that involve transition
densities with non-standard shapes. This is not surprising upon
inspecting the model structure in (\ref{stationary-mixture}), and contrasting 
it with (\ref{eqn:trans_dens_reparam}) and (\ref{eqn:weights_reparam}). 
In particular, note that the stationarity restriction forces a single
set of mixing parameters $\mu_{l}$ used to inform both
the means of the Gaussian AR mixture components and the locations of
the associated mixture weights. Moreover, the $\sigma^{2}_{l}$
control the dispersion of both the Gaussian mixture components 
and of the Gaussian densities that define the mixture weights.

\section{Extensions}
\label{sec:extensions}

The data illustrations suggest the ability of the first-order model to uncover a variety of conditional 
density shapes, and approximate well the truth contained in simulated data. However, some applications 
may require additional features in the model formulation. 

%
%

The first-order model can be extended to accommodate higher order
structure (say, based on $r$ lagged terms), where again the transition 
density, $f(z_t\mid z_{t-1},\dots,z_{t-r})$, is implied by a joint DP mixture model. 
Hence, the
transition density has a similar form to 
(\ref{eqn:trans_dens}), but now the means of the Gaussian mixture components and the mixture
weights depend on the previous $r$ states. Let superscript $y$ correspond to $Z_t$ and $x$ 
to $(Z_{t-r},\dots,Z_{t-1})$ in the 
vector $\bm{\mu}$ of length $r+1$ and the $(r+1)\times(r+1)$ matrix $\Sigma$. Under the 
reparameterization of $\Sigma$ used in the first-order case, the
Gaussian mixture kernels 
have the form $\mathrm{N}(z_{t} \mid \mu_l^y-\sum_{j=1}^r\beta_{l,(r+1,j)}(z_{t+j-r-1}-\mu_{l,j}^x),\delta_l^y)$, for 
$l=1,\dots,L$. Gibbs sampling steps are thus preserved for $\mu_l^y$ and $\delta_l^y$, as well 
as for the last row of the matrix $\beta$. However, more care is needed in devising an MCMC algorithm 
to sample $\bm{\delta}_l^x$, $\bm{\mu}_l^x$ (each a vector of length $r$) and the first $r$ rows of $\beta$, 
particularly when $r$ is of order larger than $2$ or $3$.

%
%

Turning to an application oriented extension, in population biology, the size of a wild population is often 
monitored over time. Yearly estimated biomass may be recorded for a specific species, and the trend in 
population size indicates how the species is faring, and is indicative of greater environmental conditions. 
A state-space modeling framework is suitable for such applications, since the observed biomass is not 
an exact measurement of population size. Rather, biomass is viewed as a noisy version of the underlying 
population size, and a key goal is to forecast population size in the future. 
The proposed model can be incorporated into a state-space framework, with the addition of an observation 
equation. The observations are now viewed as arising from latent unobserved states, which evolve in time 
according to the Markovian model. Denote the observed data by $(y_1,\dots,y_n)$, and the underlying 
latent states by $(z_1,\dots,z_n)$. Assume $y_t\mid z_t,\bm{\theta} \sim$ $f(y_t \mid z_t,\bm{\theta})$, for some 
parametric distribution $f(y_t \mid z_t,\bm{\theta})$, with the latent
states evolving according to the 
nonparametric mixture model for $f(z_t \mid z_{t-1})$. 
In the population dynamics example, environmental covariates may also be available. 
These can be treated as random, and modeled jointly with $y_t$ at the observation level, or incorporated 
at the state level.

The introduction of latent states is also useful in modeling ordinal time series data, as it is often 
assumed that $Y_t=j$ if and only if $Z_t\in (\gamma_{j-1},\gamma_j)$, for $j=1,\dots,C$. 
However, rather than working with a restrictive parametric distribution for the latent continuous 
responses, they can be modeled with the proposed nonparametric Markovian model.

\section{Summary}
\label{sec:conclusion}

We have proposed a modeling approach for nonstationary time series which allows for
non-standard transition densities and nonlinear autoregressions. The transition
density of the Markovian model admits a representation as a location-scale mixture of normal
densities, with means and mixture weights that depend on observations from previous time
points. This model structure arises from the conditional 
distribution induced from a Dirichlet process mixture of multivariate normals.
We have discussed methods for posterior inference and prior
specification, and illustrated the model with synthetic and real data,

including comparison with a special case of the mixture model that
ensures existence of a stationary distribution for the Markov chain.
Although the methodology has been developed and applied for directly observable
time series with first-order dependence, we have discussed possible extensions to model 
higher order Markov chains, and to expand the model structure to a state-space setting.

\section*{Acknowledgments}

The work of the first author was supported by the National Science Foundation under 
award SES 1131897. The work of the second author was supported in part by the 
National Science Foundation under awards DMS 1310438 and DMS 1407838.
The authors wish to thank two reviewers for constructive feedback and for comments 
that improved the presentation of the material in the paper.

\vspace{0.5cm}

%
%

\section*{Appendix A. The Markov chain Monte Carlo algorithm}
\label{app:MCMC}

Here, we provide the details of the MCMC method for posterior simulation from the 
nonparametric mixture model developed in Section \ref{sec:model}.

The posterior full conditional distributions for $\alpha$ and the components of vector $\psi$ 
are standard as they are assigned conditionally conjugate priors. 
Each $U_t$, $t=2,\dots,n$ is sampled from a discrete distribution on $\{1,\dots,L\}$, 
with probabilities $(\tilde{p}_{1,t},\dots,\tilde{p}_{L,t})$, where
$\tilde{p}_{l,t}\propto p_l\mathrm{N}(z_t \mid
\mu_{l}^y-\beta_{l}(z_{t-1}-\mu_{l}^x),\delta_{l}^y) \mathrm{N}(z_{t-1} \mid \mu_l^x,\delta_l^x)$, 
for $l=1,\dots,L$.

Next, consider the mixing parameters. Letting $\{U_j^*:j=1,\dots,n^*\}$ be the $n^*$ distinct values of $(U_2,\dots,U_n)$, and $M_l=|\{U_t:U_t=l\}|$, we obtain the full conditional 
\[
p(\bm{\eta}_l \mid \dots,\mathrm{data}) \propto G_0(\bm{\eta}_l \mid \bm{\psi})
\left\{
 \prod_{j=1}^{n^*}\prod_{\{t:U_t=U_j^*\}}\mathrm{N}(z_t \mid \mu_{l}^y-\beta_{l}(z_{t-1}-\mu_{l}^x),\delta_{l}^y) \right\}
\left\{  \prod_{r=1}^L \prod_{\{t:U_t=r\}}q_r(z_{t-1}) \right\}.
\]
 Therefore, if $l\in \{U_j^*\}$, $\mu_l^y$ is sampled from a normal distribution with variance $(v^y)^*=[(\nu^y)^{-1}+M_l(\delta_l^y)^{-1}]^{-1}$, and mean $(v^y)^*[(\nu^y)^{-1}m^y+(\delta_l^y)^{-1}\sum_{\{t:U_t=U_j^*\}}(z_t+\beta_l(z_{t-1}-\mu_l^x))]$. If component $l$ is empty, that is, $l\notin \{U_j^*\}$, then $\mu_l^y\sim \mathrm{N}(m^y,v^y)$. The updates for $\delta_l^y$ and $\beta_l$ also require only Gibbs sampling. If $l\in \{U_j^*\}$, then $\delta_l^y\sim \mathrm{IG}(\nu^y+0.5M_l,s^y+0.5\sum_{\{t:U_t=l\}}(z_t-\mu_l^y+\beta_l(z_{t-1}-\mu_l^x))^2)$ and $\beta_l$ is sampled from a normal with variance $c^*=[c^{-1}+(\delta_l^y)^{-1}\sum_{\{t:U_t=l\}}(z_{t-1}-\mu_l^x)^2]^{-1}$ and mean $c^*[c^{-1}\theta+(\delta_l^y)^{-1}\sum_{\{t:U_t=l\}}(z_{t-1}-\mu_l^x)(\mu_l^y-z_t)]$. 
If $l\notin \{U_j^*\}$, then we sample from $G_0$: $\delta_l^y\sim \mathrm{IG}(\nu^y,s^y)$ and 
$\beta_l\sim \mathrm{N}(\theta,c)$.

No matter the choice of $G_0$, the full conditionals for $\mu_l^x$ and $\delta_l^x$ are not proportional 
to any standard distribution, as these parameters are contained in the sum of $L$ terms in the 
denominator of $q_l(z_{t-1})$. The posterior full conditional $p(\mu_l^x \mid \dots,\mathrm{data})$, 
when $l\in\{U_j^*\}$, is given by 
\[
\mathrm{N}\left(\mu_l^x \mid m^x,v^x\right)
\prod_{\{t:U_t=l\}}\mathrm{N}\left(z_t \mid \mu_{l}^y-\beta_{l}(z_{t-1}-\mu_{l}^x),\delta_{l}^y\right)
\mathrm{N}(z_{t-1} \mid \mu_l^x,\delta_l^x)
\left(\prod_{t=2}^n \sum_{m=1}^L p_m\mathrm{N}(z_{t-1} \mid \mu_m^x,\delta_m^x)\right)^{-1}.
\]
This can be written as $p(\mu_l^x|\dots,\mathrm{data})\propto$
$\mathrm{N}(\mu_l^x \mid (m^x)^*,(v^x)^*)(\prod_{t=2}^n \sum_{m=1}^L 
p_m\mathrm{N}(z_{t-1} \mid \mu_m^x,\delta_m^x))^{-1}$, with $(v^x)^*=((v^x)^{-1}+M_l(\delta_l^x)^{-1}+M_l\beta_l^2(\delta_l^y)^{-1})$ and $(m^x)^*=(v^x)^*((v^x)^{-1}m^x+(\delta_l^x)^{-1}\sum_{\{t:U_t=l\}}z_{t-1}+(\delta_l^y)^{-1}\beta_l^2\sum_{\{t:U_t=l\}}(z_{t-1}+(z_t-\mu_l^y)/\beta_l))$. We use a random-walk Metropolis step to update $\mu_l^x$. For $l\notin\{U_j^*\}$, 
$p(\mu_l^x \mid \dots,\mathrm{data})$ is proportional to 
$\mathrm{N}(\mu_l^x \mid m^x,v^x)[\prod_{t=2}^n \sum_{m=1}^L 
p_m\mathrm{N}(z_{t-1} \mid \mu_m^x,\delta_m^x)]^{-1}$, and in this case we 
use a Metropolis-Hastings algorithm, proposing  a candidate value $\mu_l^x$ 
from the base distribution $\mathrm{N}(m^x,v^x)$.

The full conditional and sampling strategy for $\delta_l^x$ are similar to those for $\mu_l^x$. We have 
\[
p(\delta_l^x \mid \dots,\mathrm{data})\propto\mathrm{IG}(\delta_l^x \mid \nu^x,s^x)
\prod_{\{t:U_t=l\}} \mathrm{N}(z_{t-1} \mid \mu_l^x,\delta_l^x)
\left(\prod_{t=2}^n \sum_{m=1}^L p_m\mathrm{N}(z_{t-1} \mid \mu_m^x,\delta_m^x)\right)^{-1},
\]
which for an active component, is written as proportional to 
\[\mathrm{IG}\left(\delta_l^x \mid \nu^x+0.5M_l,s^x+0.5\sum_{\{t:U_t=l\}}(z_{t-1}-\mu_l^x)^2\right)\left(\prod_{t=2}^n \sum_{m=1}^L p_m\mathrm{N}(z_{t-1} \mid \mu_m^x,\delta_m^x)\right)^{-1}.\]
 For non-active components, the full conditional is $\mathrm{IG}(\delta_l^x \mid \nu^x,s^x)(\prod_{t=2}^n \sum_{m=1}^L p_m\mathrm{N}(z_{t-1} \mid \mu_m^x,\delta_m^x))^{-1}$. We use a similar strategy for sampling $\delta_l^x$ as we did with $\mu_l^x$, using a random-walk Metropolis algorithm for the active components of $\delta_l^x$, working on the log-scale and sampling $\log(\delta_l^x)$, and proposing the non-active components from $G_0(\delta_l^x)=$ $\mathrm{IG}(\nu^x,s^x)$.

We next discuss the updating scheme for the vector $\bm{p}=$ $(p_1,\dots,p_L)$, which 
poses the main challenge for posterior simulation. The full conditional for $\bm{p}$ has the form 
$$
f(\bm{p} \mid \alpha)\prod_{l=1}^L p_l^{M_l}
\left(
\prod_{t=2}^n \sum_{m=1}^L p_m\mathrm{N}(z_{t-1} \mid \mu_m^x,\delta_m^x) \right)^{-1}.
$$
In standard DP mixture models, the implied generalized Dirichlet prior for $f(\bm{p} \mid \alpha)$ combines with $\prod_{l=1}^L p_l^{M_l}$ to form another generalized Dirichlet distribution. However, in this case there is an additional term. Metropolis--Hastings algorithms with various proposal distributions were explored to sample the vector $p$,  resulting in very low acceptance rates. We instead devise an alternative sampling scheme, in which we work directly with the latent beta-distributed random variables which determine the probability vector $\bm{p}$ arising 
from the DP truncation approximation. Recall that 
$p_1=v_1$, $p_l=v_l\prod_{r=1}^{l-1}(1-v_r)$, for $l=2,\dots,L-1$, and $p_L=\prod_{r=1}^{L-1}(1-v_r)$,
where $v_1,\dots,v_{L-1}\stackrel{i.i.d.}{\sim}\mathrm{beta}(1,\alpha)$. Equivalently, let $\zeta_1,\dots,\zeta_{L-1}\stackrel{i.i.d.}{\sim}\mathrm{beta}(\alpha,1)$, and define $p_1=$ $1-\zeta_1$, 
$p_l =$ $(1-\zeta_l)\prod_{r=1}^{l-1}\zeta_r$, and $p_L=\prod_{r=1}^{L-1}\zeta_r$. 
Rather than updating directly $\bm{p}$, we work with the $\zeta_{l}$, a sample for which 
implies a particular probability vector $\bm{p}$.

The full conditional for $\zeta_l$, $l=1,\dots,L-1$, has the form
\begin{equation}
\tag{A.1}
p(\zeta_{l} \mid \dots,\mathrm{data}) \propto \mathrm{beta}\left(\zeta_{l} \mid \alpha+\sum_{r=l+1}^{L}M_{r},M_{l}+1\right)\left(\prod_{t=2}^{n}d(z_{t-1})\right)^{-1}
\end{equation}
where 
\[
d(z_{t-1})=\mathrm{N}(z_{t-1} \mid \mu_{1}^{x},\delta_{1}^{x})(1-\zeta_{1})+\sum_{l=2}^{L-1}
\mathrm{N}(z_{t-1} \mid \mu_{l}^{x},\delta_{l}^{x})(1-\zeta_{l})\prod_{s=1}^{l-1}\zeta_{s}+
\mathrm{N}(z_{t-1} \mid \mu_{L}^{x},\delta_{L}^{x})\prod_{s=1}^{L-1}\zeta_{s}.
\]
Also, let $c_{t,l}=\mathrm{N}(z_{t-1} \mid \mu_{l}^{x},\delta_{l}^{x})$, which is constant with respect 
to each $\zeta_l$. The form of the full conditional in 
(A.1) suggests the use of 
a slice sampler to update each $\zeta_l$ one at a time. The slice sampler is implemented by 
drawing auxiliary random variables $u_{t}\sim \mathrm{uniform}(0,(d(z_{t-1}))^{-1}),$ $t=2,...,n,$ 
and then sampling $\zeta_{l}\sim \mathrm{beta}(\alpha+\sum_{r=l+1}^{L}M_{r},M_{l}+1)$, but 
restricted to the set $\{\zeta_{l}:u_{t}<(d(z_{t-1}))^{-1},t=2,...,n\}$. The term $d(z_{t-1})$
can be expressed as 
%
%
$d(z_{t-1})=$ $\zeta_{l} w_{1t}+w_{0t}$, for any $l=1,...,L-1$, where 
\[w_{1t}=-c_{t,l}\prod_{s=1}^{l-1}\zeta_{s}+\left(\sum_{m=l+1}^{L-1}c_{t,m}(1-\zeta_{m})\prod_{s=1,s\neq l}^{m-1}\zeta_{s}\right)+c_{t,L}\prod_{s=1,s\neq l}^{L-1}\zeta_{s}\]
 and, if $l=1,$ $w_{0t}=c_{t,1}$, otherwise $w_{0t}=c_{t,1}(1-\zeta_{1})+\sum_{s=2}^{l-1}c_{t,s}(1-\zeta_{s})\prod_{r=1}^{s-1}\zeta_{r}+c_{t,l}\prod_{s=1}^{l-1}\zeta_{s}$. 
Then, the set $\{\zeta_{l}:d(z_{t-1})<u_{t}^{-1}\}$ is $\{\zeta_{l}:\zeta_{l}w_{1t}<u_{t}^{-1}-w_{0t}\}.$ This takes the form of $\{\zeta_{l}:\zeta_{l}<(u_{t}w_{1t})^{-1}-w_{0t}(w_{1t})^{-1}\}$ when $w_{1t}$ is positive, and has the form $\{\zeta_{l}:\zeta_{l}>(u_{t}w_{1t})^{-1}-w_{0t}(w_{1t})^{-1}\}$ otherwise. Therefore, the truncated--beta random draw for $\zeta_l$ must lie in the interval $(\max_{\{t:w_{1t}<0\}}[(u_{t}w_{1t})^{-1}-w_{0t}(w_{1t})^{-1}],\min_{\{t:w_{1t}>0\}}[(u_{t}w_{1t})^{-1}-w_{0t}(w_{1t})^{-1}])$. 
The inverse CDF random variate generation method can be used to sample from these truncated beta 
random variables. This strategy results in direct draws for the $\zeta_l$.

\section*{Appendix B. Computing posterior predictive ordinates}
\label{app:PPO}

We describe here an approach to computing one-step-ahead posterior predictive ordinates, 
$p(z_{t} \mid \bm{z}_{(t-1)})$, where $\bm{z}_{(m)}=$ $(z_{2},...,z_{m})$, for $m=2,...,n$, is 
the observed series up to time $m$. The objective is to compute $p(z_{t} \mid \bm{z}_{(t-1)})$
for any desired number of observations $z_{t}$, using the samples from the
posterior distribution given the full data vector $\bm{z}_{(n)}$.

Denote by $\bm{\Theta}=$ $(\{ \bm{\eta}_{l}: l=1,...,L \},\bm{p},\alpha,\bm{\psi})$ 
all model parameters, excluding the latent configuration variables. 
We abbreviate $f(z_t \mid z_{t-1},G)$ in (\ref{eqn:trans_dens_reparam}) to 
$f(z_t \mid z_{t-1})$, but note that, given the $\bm{\eta}_{l}$ and
$\bm{p}$, the mixture model for the transition density can be
computed at any values $z_{t}$ and $z_{t-1}$. Let $B_{(m)}$ be
the normalizing constant of the posterior distribution for $\bm{\Theta}$
given $\bm{z}_{(m)}$, and $p(\bm{\Theta})=$ 
$\{ \prod_{l=1}^{L} G_0(\bm{\eta}_l \mid \bm{\psi}) \} f(\bm{p} \mid \alpha)
p(\alpha) p(\bm{\psi})$ be the prior for $\bm{\Theta}$. Then,
\[
p(\bm{\Theta} \mid \bm{z}_{(n-1)}) = 
\frac{p(\bm{\Theta}) \prod_{t=2}^{n-1} f(z_{t} \mid z_{t-1}) }{B_{(n-1)}} =
\frac{p(\bm{\Theta}) \prod_{t=2}^{n} f(z_{t} \mid z_{t-1}) }{B_{(n-1)}\, f(z_{n} \mid z_{n-1})} =
\frac{ B_{(n)} \, p(\bm{\Theta} \mid \bm{z}_{(n)})}{ B_{(n-1)} \, f(z_{n} \mid z_{n-1})}
\]
and therefore $p(z_{n} \mid \bm{z}_{(n-1)}) =$ 
$\int f(z_{n} \mid z_{n-1}) p(\bm{\Theta} \mid \bm{z}_{(n-1)}) \, \text{d}\bm{\Theta}=$
$B_{(n)}/B_{(n-1)}$. In addition, 
$\int \{ f(z_{n} \mid z_{n-1}) \}^{-1} p(\bm{\Theta} \mid \bm{z}_{(n)}) \, \text{d}\bm{\Theta}=$
$B_{(n-1)}/B_{(n)}$, and thus 
\[
\tag{B.1}
p(z_{n} \mid \bm{z}_{(n-1)}) = \left( 
\int \{ f(z_{n} \mid z_{n-1}) \}^{-1} p(\bm{\Theta} \mid \bm{z}_{(n)}) \, \text{d}\bm{\Theta}
\right)^{-1}.
\]
Similarly, $p(\bm{\Theta} \mid \bm{z}_{(n-2)})=$ 
$\{ B_{(n)} p(\bm{\Theta} \mid \bm{z}_{(n)}) \}/\{ B_{(n-2)} f(z_{n} \mid z_{n-1}) f(z_{n-1} \mid z_{n-2}) \}$.
Hence, $p(z_{n-1} \mid \bm{z}_{(n-2)})=$
$\int f(z_{n-1} \mid z_{n-2}) p(\bm{\Theta} \mid \bm{z}_{(n-2)}) \, \text{d}\bm{\Theta}=$
$\frac{ B_{(n)} }{ B_{(n-2)} } 
\int \{ f(z_{n} \mid z_{n-1}) \}^{-1} p(\bm{\Theta} \mid \bm{z}_{(n)}) \, \text{d}\bm{\Theta}$.
Then, observing that 
$\int \{ f(z_{n} \mid z_{n-1}) f(z_{n-1} \mid z_{n-2}) \}^{-1} p(\bm{\Theta} \mid \bm{z}_{(n)}) \, 
\text{d}\bm{\Theta}=$ $B_{(n-2)}/B_{(n)}$, we obtain an expression for $p(z_{n-1} \mid \bm{z}_{(n-2)})$
that involves the product of the two integrals above. Extending the
derivation for $p(z_{n-1} \mid \bm{z}_{(n-2)})$, we obtain 
\[
p(z_{t} \mid \bm{z}_{(t-1)}) = \left( 
\int \left\{ \prod_{s=t}^{n} f(z_{s} \mid z_{s-1}) \right\}^{-1} 
p(\bm{\Theta} \mid \bm{z}_{(n)}) \, \text{d}\bm{\Theta}
\right)^{-1} \left( 
\int \left\{ \prod_{s=t+1}^{n} f(z_{s} \mid z_{s-1}) \right\}^{-1} 
p(\bm{\Theta} \mid \bm{z}_{(n)}) \, \text{d}\bm{\Theta}
\right)
\]
for any $t=3,...,n-1$, with the expression for $t=n$ given in (B.1). These 
expressions allow us to estimate any posterior predictive ordinate 
$p(z_{t} \mid \bm{z}_{(t-1)})$, using Monte Carlo integration
based on the samples from $p(\bm{\Theta} \mid \bm{z}_{(n)})$.

\vspace{0.5cm}

\singlespacing

\bibliographystyle{asa}

\bibliography{DPM_MC_bib}

\begin{thebibliography}{29}
\newcommand{\enquote}[1]{``#1''}
\expandafter\ifx\csname natexlab\endcsname\relax\def\natexlab#1{#1}\fi

\bibitem[{Antoniano-Villalobos and Walker(2016)}]{antoniano}
Antoniano-Villalobos, I. and Walker, S.~G. (2016), \enquote{A nonparametric
  model for stationary time series,} \textit{Journal of Time Series Analysis},
  37, 126--142.

\bibitem[{Azzalini(1985)}]{azzalini}
Azzalini, A. (1985), \enquote{A class of distributions which includes the
  normal ones,} \textit{Scandinavian Journal of Statistics}, 12, 171--178.

\bibitem[{Caron et~al.(2008)Caron, Davy, Doucet, Duflos, and Vanheeghe}]{caron}
Caron, F., Davy, M., Doucet, A., Duflos, E., and Vanheeghe, P. (2008),
  \enquote{Bayesian Inference for linear dynamic models with Dirichlet process
  mixtures,} \textit{IEEE Transactions on Signal Processing}, 56, 71--84.

\bibitem[{Carvalho and Tanner(2005)}]{carvalho}
Carvalho, A. and Tanner, M. (2005), \enquote{Modeling nonlinear time series
  with local mixtures of generalized linear models,} \textit{Canadian Journal
  of Statistics}, 33, 97--113.

\bibitem[{Carvalho and Tanner(2006)}]{carvalho2006}
--- (2006), \enquote{Modeling nonlinearities with mixtures of experts time
  series models,} \textit{International Journal of Mathematics and Mathematical
  Sciences}, 2006.

\bibitem[{Connor and Mosimann(1969)}]{connor}
Connor, R. and Mosimann, J. (1969), \enquote{Concepts of independence for
  proportions with a generalization of the {D}irichlet distribution,}
  \textit{Journal of the American Statistical Association}, 64, 194--206.

\bibitem[{Daniels and Pourahmadi(2002)}]{daniels}
Daniels, M. and Pourahmadi, M. (2002), \enquote{Bayesian analysis of covariance
  matrices and dynamic models for longitudinal data,} \textit{Biometrika}, 89,
  553--566.

\bibitem[{DeYoreo and Kottas(2015)}]{deyoreokottas}
DeYoreo, M. and Kottas, A. (2015), \enquote{A fully nonparametric modeling
  approach to binary regression,} \textit{Bayesian Analysis}, 10, 821--847.

\bibitem[{Di{ }Lucca et~al.(2013)Di{ }Lucca, Guglielmi, M{\"u}ller, and
  Quintana}]{dilucca}
Di{ }Lucca, M., Guglielmi, A., M{\"u}ller, P., and Quintana, F. (2013),
  \enquote{A simple class of {B}ayesian autoregression models,}
  \textit{Bayesian Analysis}, 8, 63--88.

\bibitem[{Ferguson(1973)}]{ferguson}
Ferguson, T. (1973), \enquote{A {B}ayesian analysis of some nonparametric
  problems,} \textit{The Annals of Statistics}, 1, 209--230.

\bibitem[{Fox et~al.(2011)Fox, Sudderth, Jordan, and Willsky}]{fox}
Fox, E., Sudderth, E., Jordan, M., and Willsky, A. (2011), \enquote{Bayesian
  nonparametric inference for switching dynamic linear models,} \textit{IEEE
  Transactions on Signal Processing}, 59, 1569--1585.

\bibitem[{Fr{\"u}wirth-Schnatter(2006)}]{fruwirth}
Fr{\"u}wirth-Schnatter, S. (2006), \textit{Finite Mixture and Markov Switching
  Models}, Springer.

\bibitem[{Geweke and Terui(1993)}]{geweke}
Geweke, J. and Terui, N. (1993), \enquote{Bayesian threshold autoregressive
  models for nonlinear time series,} \textit{Journal of Time Series Analysis},
  14, 441--454.

\bibitem[{Ishwaran and James(2001)}]{ishjames}
Ishwaran, H. and James, L. (2001), \enquote{Gibbs sampling methods for
  stick-breaking priors,} \textit{Journal of the American Statistical
  Association}, 96, 161--173.

\bibitem[{Juang and Rabiner(1985)}]{juangrabiner}
Juang and Rabiner (1985), \enquote{Mixture autoregressive hidden Markov models
  for speech signals,} \textit{IEEE Transactions and Acoustic, Speech, and
  Signal Processing}, 1404--1413.

\bibitem[{Lau and So(2008)}]{lauso}
Lau, J. and So, M. (2008), \enquote{Bayesian mixture of autoregressive models,}
  \textit{Computational Statistics and Data Analysis}, 53, 38--60.

\bibitem[{MacEachern(2000)}]{maceachern}
MacEachern, S. (2000), \enquote{Dependent Dirichlet processes,} Tech. rep., The
  Ohio State University, Department of Statistics.

\bibitem[{Martinez-Ovando and Walker(2011)}]{martinez}
Martinez-Ovando, J. and Walker, S.~G. (2011), \enquote{Time-series modeling,
  stationarity, and {B}ayesian nonparametric methods,} Tech. rep., Banco de
  Mexico.

\bibitem[{Mena and Walker(2005)}]{mena}
Mena, R. and Walker, S. (2005), \enquote{Stationary autoregressive models via a
  Bayesian nonparametric approach,} \textit{Journal of Time Series Analysis},
  26, 789--805.

\bibitem[{M{\"u}ller et~al.(1997)M{\"u}ller, West, and
  MacEachern}]{mullerwestmac}
M{\"u}ller, P., West, M., and MacEachern, S. (1997), \enquote{Bayesian models
  for nonlinear autoregressions,} \textit{Journal of Time Series Analysis}, 18,
  593--614.

\bibitem[{Sethuraman(1994)}]{sethuraman}
Sethuraman, J. (1994), \enquote{A constructive definition of {D}irichlet
  priors,} \textit{Statistica Sinica}, 4, 639--650.

\bibitem[{Tang and Ghosal(2007{\natexlab{a}})}]{tang:cs}
Tang, Y. and Ghosal, S. (2007{\natexlab{a}}), \enquote{A consistent
  nonparametric {B}ayesian procedure for estimating autoregressive conditional
  densities,} \textit{Computational Statistics and Data Analysis}, 51,
  4424--4437.

\bibitem[{Tang and Ghosal(2007{\natexlab{b}})}]{tang:planning}
--- (2007{\natexlab{b}}), \enquote{Posterior consistency of Dirichlet mixtures
  for estimating a transition density,} \textit{Journal of Statistical Planning
  and Inference}, 137, 1711--1726.

\bibitem[{Tong(1987)}]{tong}
Tong, H. (1987), \enquote{On a Threshold Model,} in \textit{Pattern Recognition
  and Signal Processing}, ed. Chen, C., Amsterdam: Sijhoff and Nordhoff.

\bibitem[{Tong(1990)}]{tong1990}
--- (1990), \textit{Non-Linear Time Series: A Dynamical System Approach},
  Oxford University Press.

\bibitem[{Webb and Forster(2008)}]{webb}
Webb, E. and Forster, J. (2008), \enquote{Bayesian model determination for
  multivariate ordinal and binary data,} \textit{Computational Statistics and
  Data Analysis}, 52, 2632--2649.

\bibitem[{West and Harrison(1999)}]{westharrison}
West, M. and Harrison, J. (1999), \textit{Bayesian Forecasting and Dynamic
  Models}, New York: Springer.

\bibitem[{Wong and Li(2000)}]{wongli}
Wong, C.~S. and Li, W.~K. (2000), \enquote{On a mixture autoregressive model,}
  \textit{Journal of the Royal Statistical Society, Series B}, 62, 95--115.

\bibitem[{Wood et~al.(2011)Wood, Rosen, and Kohn}]{wood}
Wood, S., Rosen, O., and Kohn, R. (2011), \enquote{Bayesian mixtures of
  autoregressive models,} \textit{Journal of Computational and Graphical
  Statistics}, 20, 174--195.

\end{thebibliography}

\end{document}